\documentclass[amsmath,aps,showpacs,a4paper,10pt]{revtex4}

 \usepackage{epsf}
 \usepackage{graphicx}    

 \usepackage{rotating}    

\begin{document}

 \newcommand{\be}[1]{\begin{equation}\label{#1}}
 \newcommand{\ee}{\end{equation}}
 \newcommand{\bea}{\begin{eqnarray}}
 \newcommand{\eea}{\end{eqnarray}}
 \def\disp{\displaystyle}

 \def\ssrm{\scriptscriptstyle \rm}   

 \begin{titlepage}

 \begin{flushright}
 \end{flushright}

 \title{\Large \bf Guide for the Hubble Tension: Mid-Time Solutions
 \\ and Generalized PAge Parameterizations}

 \author{Jing-Yi~Jia\,$^{a,}$\footnote{email
 address:\ jjy@bit.edu.cn}\,,
 Da-Chun~Qiang\,$^b$\,,
 Hao~Wei\,$^{a,}$\footnote{Corresponding author;\ email
 address:\ haowei@bit.edu.cn}\vspace{2.4mm}}
 \affiliation{$^{a)\,}$School of Physics, Beijing
 Institute of Technology, Beijing 100081, China\vspace{2mm}\\
 $^{b)\,}$Institute for Gravitational Wave Astronomy, Henan Academy of
 Sciences, Zhengzhou 450046, Henan, China}

 \begin{abstract}\vspace{1cm}
 \centerline{\bf ABSTRACT}\vspace{2mm}
 In cosmology, the Hubble tension has become one of the most serious
 challenges in the last decade. Assuming that all the observational
 data are right, one of the ways out is to modify the standard
 $\Lambda$CDM cosmology.~In the literature, the theoretical
 modifications are mainly made in the early and the late/local
 universes.~But on both sides, some no-go arguments have been proposed
 recently.~In particular, the general no-go arguments for the late-time
 modifications are mainly based on the PAge parameterization.~In the
 present work, we note that there are some flaws in the (original) PAge
 parameterization, and then propose various generalized PAge (GPAge)
 parameterizations, which can be extended to $z>z_{\rm CMB}$ and hence
 the cosmic microwave background (CMB) can be taken into account, while
 they are also more accurate at low redshifts.~With these GPAge
 parameterizations, we revisit the Hubble tension by using not only the
 late-time observations but also the observational data of CMB.~We find
 that the Hubble tension could be alleviated (or even resolved)
 in GPAges significantly different from the standard $\Lambda$CDM cosmology,
 while they are overwhelmingly preferred over $\Lambda$CDM by
 the observational data in terms of all the information
 criteria.~Surprisingly, we find that the new physics in the mid-time
 might be the key to resolve the Hubble tension.
 \end{abstract}

 \pacs{98.80.-k, 98.80.Es, 95.36.+x, 04.50.Kd}

 \maketitle

 \end{titlepage}

 \renewcommand{\baselinestretch}{1.0}


\section{Introduction}\label{sec1}

In cosmology, the Hubble tension has been one of the most serious challenges
 in the last decade~\cite{DiValentino:2025sru,Abdalla:2022yfr,
 Perivolaropoulos:2021jda,DiValentino:2021izs,Rong-Gen:2023dcz,
 H0DN:2025lyy}. Assuming the $\Lambda$CDM model, the Hubble constant
 inferred from the final Planck measurements (Planck 2018) of
 the cosmic microwave background (CMB) is given
 by $H_0=67.36\pm 0.54\;{\rm km/s/Mpc}$~\cite{Planck:2018vyg}.~On the
 other hand, based on the Cepheid/Type Ia supernova (SNIa) distance
 ladder, the local determination of $H_0$ from the Hubble Space
 Telescope and the SH0ES team is given by $H_0=73.04\pm
 1.04\;{\rm km/s/Mpc}$~\cite{Riess:2021jrx}. Clearly, the Hubble constants
 inferred from the late- and the early-universes are in a great tension
 $>5\sigma$. Recently, the Hubble tension has become $>6\sim
 7\sigma$~\cite{Riess:2024vfa,Riess:2025chq,H0DN:2025lyy}.~Nowadays,
 it casts a serious shadow over the standard cosmology.

One of the ways out is to carefully check the observational data, which
 might have some unresolved systematic errors.~Many efforts have been
 made on both sides of CMB and SNIa/Cepheid in the literature, and no
 convincing evidences have been found to date~\cite{DiValentino:2025sru,
 Abdalla:2022yfr,Perivolaropoulos:2021jda,DiValentino:2021izs,
 Rong-Gen:2023dcz}.

Assuming that all the observational data are right, the other way out
 is to modify the standard $\Lambda$CDM cosmology.~Of course, various
 trivial extensions to the base-$\Lambda$CDM model do not
 work~\cite{DiValentino:2025sru,Abdalla:2022yfr,
 Perivolaropoulos:2021jda,DiValentino:2021izs,Rong-Gen:2023dcz}.~The
 Hubble tension invokes non-trivial modifications in the early, middle,
 late or local universes.

In the literature, the theoretical modifications are mainly made in the
 early and the late/local universes. For example, one of the leading
 models to resolve the Hubble tension in the early universe is
 the well-known early dark energy (EDE) model~\cite{Poulin:2018cxd}
 and its variants (see e.g.~\cite{Kamionkowski:2022pkx,Poulin:2023lkg,
 McDonough:2023qcu} for reviews).~In particular, the Anti-de Sitter
 (AdS) EDE model~\cite{Ye:2020btb,Ye:2020oix,Jiang:2021bab,Wang:2025dtk}
 is fairly interesting.~Also, the dark radiation is one of the
 well-motivated and straightforward extensions to resolve the Hubble
 tension in the early universe (see
 e.g.~\cite{Gariazzo:2023hch} and~\cite{DiValentino:2025sru,Abdalla:2022yfr,
 Perivolaropoulos:2021jda,DiValentino:2021izs,Rong-Gen:2023dcz} for
 reviews).~However, there is a well-known no-go
 argument~\cite{Jedamzik:2020zmd} for the early-time solutions (see also
 e.g.~\cite{DiValentino:2025sru,Abdalla:2022yfr,Perivolaropoulos:2021jda,
 DiValentino:2021izs,Rong-Gen:2023dcz}).~The debate has not been settled
 to date.

On the other hand, the models to resolve the Hubble tension in
 the late/local universes include various exotic (late) dark energy models,
 interacting dark energy models, dark energy models with transitions, local
 void models, and so on (see e.g.~\cite{DiValentino:2025sru,
 Abdalla:2022yfr,Perivolaropoulos:2021jda,DiValentino:2021izs,
 Rong-Gen:2023dcz} for reviews).~In particular, the phantom-like
 dark energy (PDE) transition model (see e.g.~\cite{Mortonson:2009qq,
 Efstathiou:2021ocp}) and the Chameleon dark energy model (see
 e.g.~\cite{Cai:2021wgv} and \cite{Khoury:2003aq,Khoury:2003rn,
 Wei:2004rw,Wei:2021xek}) are fairly interesting.~Note that there are
 also no-go arguments (e.g.~\cite{Benevento:2020fev,Camarena:2021jlr,
 Lemos:2018smw,Efstathiou:2021ocp,Zhang:2020uan,Cai:2021weh,Cai:2022dkh,
 Huang:2024erq}) for the late-time solutions. Nevertheless, the
 relevant works on this topic are still active.

In particular, the most general no-go argument for the
 late-time solutions of the Hubble tension comes from~\cite{Cai:2021weh,
 Cai:2022dkh,Huang:2024erq}.~To be as general as possible, one should be
 completely ignorant of the cosmology, especially the components in the
 universe and the gravity theory.~So, all the $w(z)$ parameterizations
 of dark energy are not suitable, since we pretend to even do not know
 whether the universe contains dark energy or dark matter.~To this end, the
 well-known cosmography (see e.g.~\cite{Visser:2003vq,Weinberg:2008zzc,
 Yin:2019rgm,Zhou:2016nik}) might be suitable, which is the Taylor expansion
 of the luminosity distance $d_L$ with respect to redshift $z$
 or $y$-shift $y=1-a=z/(1+z)$.~However, as shown in~\cite{Cai:2021weh,
 Cai:2022dkh}, both the cosmographies with respect to $z$ and $y$ even
 up to the fifth order cannot faithfully approximate the model
 they claim to parameterize at redshift $z\gtrsim 1$, even for
 the simplest $\Lambda$CDM model.~Therefore,
 in~\cite{Cai:2021weh,Cai:2022dkh,Huang:2024erq}, the so-called PAge
 parameterization proposed in~\cite{Huang:2020mub,Luo:2020ufj} was used
 instead, namely
 \be{eq1}
 Ht=\frac{2}{3}+\left(H_0 t_0 -\frac{2}{3}\left(1+\eta\right)\right)
 \left(\frac{t}{t_0}\right)+\frac{2}{3}\,\eta\left(\frac{t}{t_0}\right)^2\,,
 \ee
 which is in fact the Taylor expansion of $Ht$ with respect to $\tau=t/t_0$
 around $\tau=0$ up to the second order, where $H$, $t$ and $t_0$ are
 the Hubble parameter, the cosmic time, and the current age of the universe,
 respectively.~Note that the factor $2/3$ comes from requiring $Ht\to
 2/3$ as $t\to 0$~\cite{Huang:2020mub,Luo:2020ufj}, because $Ht=2/3$ in
 a matter-dominated universe as well known.~Using this PAge
 parameterization, we are completely ignorant of the components
 in the universe and the gravity theory.~It is fully model-independent.~In
 fact, the general no-go arguments in~\cite{Cai:2021weh,Cai:2022dkh,
 Huang:2024erq} are mainly based on this PAge parameterization.

However, we note that there are some flaws in the PAge parameterization
 given by Eq.~(\ref{eq1}).~First, let us roughly estimate its range of
 validity by using a simple $\Lambda$CDM model, in which $E^2=(H/H_0)^2=
 \Omega_{m0}(1+z)^3+\Omega_{r0}(1+z)^4+(1-\Omega_{m0}-\Omega_{r0})$,
 where $\Omega_{r0}=\Omega_{m0}/(1+z_{\rm eq})$ while $z_{\rm eq}\sim 3600$
 for $\Omega_{m0}\sim 0.3$ and $h=H_0/(100\;{\rm km/s/Mpc})\sim 0.7$.~We
 roughly require $\rho_r\lesssim\rho_m/120$ for convenience to estimate
 the redshift $z_{\rm MD}$ where the matter dominates.~It is equivalent
 to $\Omega_{r0}(1+z_{\rm MD})^4\lesssim\Omega_{m0}(1+z_{\rm MD})^3/120$,
 and hence $z_{\rm MD}\lesssim 30$.~So, the PAge parameterization given
 by Eq.~(\ref{eq1}) holds only after the matter-dominated redshift,
 namely $z<z_{\rm MD}\lesssim 30$.~Although this is just a rough estimation,
 it is robust to say that the PAge parameterization in Eq.~(\ref{eq1})
 is valid only at redshifts $z<{\cal O}(10)$.~Second, $z_{\rm CMB}\sim 1090$
 as well known, and hence $\rho_r\sim 0.3\,\rho_m$ at $z_{\rm CMB}$,
 where is not matter- or radiation-dominated.~The PAge parameterization
 in Eq.~(\ref{eq1}) cannot be applied to CMB.~This is the deep reason
 to only use the PAge parameterization~(\ref{eq1}) with the late-time
 observational data in the literature (especially~\cite{Cai:2021weh,
 Cai:2022dkh,Huang:2024erq}).~However, the Hubble tension is mainly
 between CMB and SNIa/Cepheids.~The no-go arguments based on the PAge
 parameterization~(\ref{eq1}) in~\cite{Cai:2021weh,Cai:2022dkh,
 Huang:2024erq} are not so solid without CMB.~Finally, the Taylor expansion
 of $Ht$ with respect to $\tau=t/t_0$ around $\tau=0$ in Eq.~(\ref{eq1})
 is only valid for $\tau\ll 1$, namely $t\ll t_0$.~At fairly low redshifts,
 $\tau$ is close to $1$ when $t$ is close to $t_0$, the higher orders
 ${\cal O}(\tau^3)$, ${\cal O}(\tau^4)\dots$ cannot be dropped
 in the Taylor expansion.~In this case, it is more reasonable to instead
 consider the Taylor expansion of $Ht$ with respect to $\tau=t/t_0$
 around $\tau=1$ (see below).~These flaws could be fixed by generalizing
 the PAge parameterization.

In the present work, we are interested in alleviating the Hubble tension
 with various generalized PAge parameterizations, which can be extended
 to redshifts $z>z_{\rm CMB}$ and hence the CMB observations can be
 taken into account, while the Taylor expansion of $Ht$ could be more
 accurate at fairly low redshifts. Surprisingly, we find that the new
 physics in the mid-time (and the early time) might be one of the keys
 to resolve the Hubble tension.

The rest of this paper is organized as follows.~In Sec.~\ref{sec2},
 we propose various generalized PAge (GPAge) parameterizations
 in detail.~In Sec.~\ref{sec3}, we briefly introduce the observational
 data and the information criteria for model comparison.~In
 Sec.~\ref{sec4}, the generalized PAge parameterizations are confronted
 with the observational data. In Sec.~\ref{sec5}, we discuss
 the physical implications and find the new physics to resolve
 the Hubble tension.~Finally, some brief concluding remarks are given in
 Sec.~\ref{sec6}.


\section{Generalized PAge parameterizations}\label{sec2}


\subsection{Single generalized PAge (SGPAge)}\label{sec2a}

As mentioned above, the PAge parameterization given by Eq.~(\ref{eq1})
 holds only at redshifts $z<{\cal O}(10)$, mainly due to the requirement
 $Ht\to 2/3$ as $t\to 0$ in a matter-dominated universe~\cite{Huang:2020mub,
 Luo:2020ufj,Huang:2024erq}.~In order to take CMB into account, the simplest
 way to extend the PAge parameterization up to redshifts $z>z_{\rm CMB}$
 is to consider the Taylor expansion of $Ht$ with respect
 to $\tau=t/t_0$ around $\tau=0$ up to the second order,
 \be{eq2}
 Ht=\alpha_0+\alpha_1\tau+\alpha_2\,\tau^2\,,
 \ee
 where $\alpha_i$ are all constant model parameters, and $H$, $t$, $t_0$ are
 the Hubble parameter, the cosmic time, the current age of the universe,
 respectively.~$Ht\to\alpha_0$ as $t\to 0$.~Noting that $Ht=2/3$ and
 $1/2$ respectively in the matter- and radiation-dominated universes as well
 known, the single generalized PAge (SGPAge) parameterization given by
 Eq.~(\ref{eq2}) reduces to the original PAge given by Eq.~(\ref{eq1})
 if $\alpha_0=2/3$, and it could be extended to the radiation-dominated
 epoch at $z_{\rm RD}$ if $\alpha_0=1/2$.~However, as mentioned above,
 the universe is not matter- or radiation-dominated at $z_{\rm CMB}\sim
 1090$, and hence $\alpha_0$ is not $2/3$ or $1/2$ for the general case
 of $z_{\rm CMB}<z<z_{\rm eq}<z_{\rm RD}$, where $z_{\rm eq}$ is the
 redshift of radiation-matter equality.~In addition, $\alpha_0$ could be
 any unknown value at redshifts $z_{\rm RD}<z<z_{\rm inflation}$.~Thus,
 we let $\alpha_0$ be a free positive model parameter in this work,
 which characterizes the validity range of GPAge.

Following e.g.~\cite{Huang:2020mub,Luo:2020ufj,Huang:2024erq,
 Cai:2021weh,Cai:2022dkh}, letting $\alpha_1=p_{\rm age}-\alpha_0\left(1
 +\eta\right)$ and $\alpha_2=\alpha_0\eta$, we recast Eq.~(\ref{eq2}) as
 \be{eq3}
 E=\frac{H}{H_0}=1+\frac{\alpha_0}{p_{\rm age}}\left(1-\eta\tau\right)
 \left(\frac{1}{\tau}-1\right)\,,
 \ee
 where $p_{\rm age}=H_0 t_0$ and $H_0=H(z=0)$ is the Hubble constant.~The
 new free model parameters $p_{\rm age}$ and $\eta$ play the roles in place
 of $\alpha_1$ and $\alpha_2$.~The dimensionless Hubble parameter $E\equiv
 H/H_0$ is a function of the normalized (dimensionless) cosmic time $\tau$
 in Eq.~(\ref{eq3}).~In order to confront SGPAge with the observational
 data, one should relate $\tau$ to redshift $z$.~Noting
 $E=(d\ln a/d\tau)/p_{\rm age}$, where $a=(1+z)^{-1}$ is the scale factor,
 Eq.~(\ref{eq3}) is in fact a differential equation of redshift $z$ with
 respect to $\tau$.~Solving this differential equation with the initial
 condition $\tau=1$ (or equivalently $t=t_0$) at $z=0$, we have
 \be{eq4}
 -\ln (1+z)=\alpha_0\ln\tau+\frac{\alpha_0}{2}\left(1-\tau\right)\left[
 \left(1-\tau\right)\eta+2-\frac{2}{\alpha_0}\,p_{\rm age}\right]\,.
 \ee
 One can numerically obtain the positive $\tau(z)$ as a function of
 redshift $z$ from Eq.~(\ref{eq4}), and then $E(z)$ as well
 as $H(z)=H_0 E(z)$ are on hand by using Eq.~(\ref{eq3}).


 \begin{center}
 \begin{figure}[tb]
 \centering
 \vspace{-7mm}  
 \includegraphics[width=0.75\textwidth]{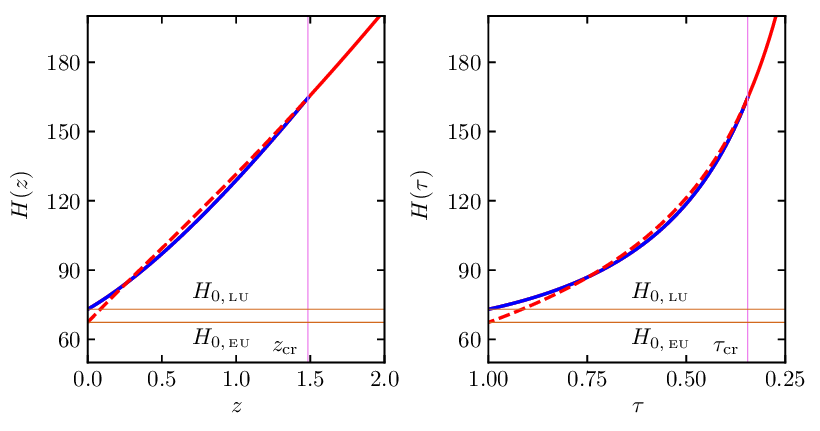}
 \vspace{-3mm}  
 \caption{\label{fig1} The Hubble parameter $H$ as functions of redshift
 $z$ (left) and $\tau$ (right) for TGPAge-Fix with the best-fit model
 parameters from the observational data SN+CMB+FAP+CC.~In particular,
 $H_{\ssrm EU}$ (red) and $H_{\ssrm LU}$ (blue) are plotted as solid
 lines, while the {\em extension lines} of $H_{\ssrm EU}$ are plotted
 as dashed red lines.~$z_{\rm cr}$, $\tau_{\rm cr}$, $H_{0,\,\ssrm EU}$
 and $H_{0,\,\ssrm LU}$ are also indicated.~This is an illustration for
 the physical meaning of $H_{0,\,\ssrm EU}$.~See Sec.~\ref{sec2b} for
 details.}
 \end{figure}
 \end{center}


\vspace{-10mm}  


\subsection{Two-segment generalized PAge (TGPAge)}\label{sec2b}

Although SGPAge can be extended to $z>z_{\rm CMB}$ and hence CMB can be
 taken into account, it is still not enough accurate at low redshifts
 where $\tau$ is close to $1$, as mentioned in Sec.~\ref{sec1}.~The simplest
 way out is to instead consider the Taylor expansion of $Ht$ with respect to
 $\tau$ around $\tau=1$ (rather than $\tau=0$) at low redshifts.~So,
 we consider a two-segment generalized PAge (TGPAge) parameterization,
 \be{eq5}
 Ht=
 \begin{cases}\\[-5mm]
 \;H_{\ssrm EU}\,t=\alpha_0+\alpha_1\tau+\alpha_2\,\tau^2
 \quad & {\rm if}\quad z\geq z_{\rm cr}\,,\\[2mm]
 \;H_{\ssrm LU}\,t=\beta_0+\beta_1\left(1-\tau\right)+\beta_2\left(1-
 \tau\right)^2 \quad & {\rm if}\quad z\leq z_{\rm cr}\,,
 \end{cases}
 \ee
 where $\alpha_i$, $\beta_i$ are all constant model parameters, and
 these two segments match at the critical redshift~$z_{\rm cr}$. The
 subscripts ``\;EU\,'' and ``\;LU\,'' denote ``\;early universe\,'' and
 ``\;late universe\,'', respectively. It is worth noting that one might
 recast the second row as $H_{\ssrm LU}\,t=\left(\beta_0+\beta_1+\beta_2
 \right)-\left(\beta_1+2\beta_2\right)\tau+\beta_2\,\tau^2$,
 which shares the same form with the first row.~However, the higher
 orders ${\cal O}(\tau^3)$, ${\cal O}(\tau^4)\dots$ cannot be dropped
 in the Taylor expansion around $\tau=0$ in the first row if
 one extrapolates it to fairly low redshifts $z\leq z_{\rm cr}$, and
 they should be incorporated into the lower orders {\em as corrections}.~So,
 it is equivalent to changing the coefficients of the lower orders, if
 one truncates the Taylor expansion up to the second order.~Through the
 above two different approaches to understanding, $H_{\ssrm LU}\,t$
 is more accurate at low redshifts $z\leq z_{\rm cr}$.

Letting $\alpha_1=p_{\rm age,\,{\ssrm EU}}-\alpha_0\left(1+
 \eta_{\ssrm EU}\right)$ and $\alpha_2=\alpha_0\eta_{\ssrm
 EU}$, we recast $H_{\ssrm EU}\,t$ in Eq.~(\ref{eq5}) as
 \be{eq6}
 E_{\ssrm EU}=\frac{H_{\ssrm EU}}{H_{0,\,{\ssrm EU}}}=
 1+\frac{\alpha_0}{p_{\rm age,\,{\ssrm EU}}}
 \left(1-\eta_{\ssrm EU}\,\tau\right)\left(\frac{1}{\tau}-1\right)\,,
 \ee
 where $p_{\rm age,\,{\ssrm EU}}=H_{0,\,{\ssrm EU}}\,t_0$.~Noting that
 in principle $\tau$ cannot be $1$ (or equivalently $t$ cannot
 be $t_0$) in $H_{\ssrm EU}\,t$ since its redshift $z$ cannot be $0$,
 $H_{0,\,{\ssrm EU}}$ should be regarded as ``\;the {\em
 inferred} Hubble constant from the early universe\,'', namely the intercept
 of the {\em extension line} of $H_{\ssrm EU}(z)$ at $z=0$ (or equivalently
 $H_{\ssrm EU}(\tau)$ at $\tau=1$).~See Fig.~\ref{fig1} for an
 illustration.~In some sense, $H_{0,\,{\ssrm EU}}$ corresponds to the
 Hubble constant inferred from CMB, and it is not the actual Hubble constant
 $H_0$ directly measured at $z=0$ by definition.~Noting $E_{\ssrm EU}=
 (d\ln a/d\tau)/p_{\rm age,\,{\ssrm EU}}$, Eq.~(\ref{eq6}) is in fact a
 differential equation of redshift $z$ with respect to $\tau$.~Solving
 this differential equation with the initial condition $\tau=1$
 (or equivalently $t=t_0$) at $z=0$, we have
 \be{eq7}
 -\ln (1+z)=\alpha_0\ln\tau+\frac{\alpha_0}{2}\left(1-\tau\right)\left[
 \left(1-\tau\right)\eta_{\ssrm EU}+2-\frac{2}{\alpha_0}\,
 p_{\rm age,\,{\ssrm EU}}\right]\,.
 \ee
 Note that the above initial condition takes the same considerations as
 for $H_{0,\,{\ssrm EU}}$ below Eq.~(\ref{eq6}).~One can numerically
 obtain the positive $\tau(z)$ as a function of redshift $z$
 from Eq.~(\ref{eq7}), and then $E_{\ssrm EU}(z)$ as well as
 $H_{\ssrm EU}(z)=H_{0,\,{\ssrm EU}}E_{\ssrm EU}(z)$ are on hand by
 using Eq.~(\ref{eq6}).


 \begin{table}[tb]
 \renewcommand{\arraystretch}{1.54}
 \begin{center}
 \vspace{-5mm}  
 \begin{tabular}{llc} \hline\hline
 Tracer & $z_{\rm eff}$ & $F_{\rm AP}$ \\ \hline
 LRG1 & $0.51$ & $0.621489\pm 0.017109$ \\
 LRG2 & $0.706$ & $0.891821\pm 0.020776$ \\
 LRG3+ELG1 \hspace{8mm} & $0.934$ \hspace{8mm} & $1.223005\pm 0.019160$ \\
 ELG2 & $1.321$ & $1.947010\pm 0.045143$ \\
 QSO & $1.484$ & $2.380581\pm 0.135882$ \\
 Ly$\alpha$ & $2.33$ & $4.517033\pm 0.096939$ \\
 \hline\hline
 \end{tabular}
 \end{center}
 \vspace{-1mm}  
 \caption{\label{tab1} The $F_{\rm AP}$ data from DESI BAO DR2
 observation.~See Sec.~\ref{sec3} for details.}
 \end{table}


Letting $\beta_0=p_{\rm age,\,{\ssrm LU}}$, $\beta_1=\xi_0
 \left(1-\eta_{\ssrm LU}\right)-p_{\rm age,\,{\ssrm LU}}$ and
 $\beta_2=\xi_0\eta_{\ssrm LU}$, we recast $H_{\ssrm LU}\,t$ in
 Eq.~(\ref{eq5}) as
 \be{eq8}
 E_{\ssrm LU}=\frac{H_{\ssrm LU}}{H_{0,\,{\ssrm LU}}}=
 1+\frac{\xi_0}{p_{\rm age,\,{\ssrm LU}}}\left(1-
 \eta_{\ssrm LU}\,\tau\right)\left(\frac{1}{\tau}-1\right)\,,
 \ee
 where $p_{\rm age,\,{\ssrm LU}}=H_{0,\,{\ssrm LU}}\,t_0$, and
 $H_{0,\,{\ssrm LU}}=H_{\ssrm LU}(\tau=1)=H_{\ssrm LU}(z=0)$ is the
 actual Hubble constant $H_0$ directly measured at $z=0$ by
 definition, which corresponds to the Hubble constant measured
 by SH0ES with nearby Cepheids/SNIa.~Similarly, noting $E_{\ssrm LU}=
 (d\ln a/d\tau)/p_{\rm age,\,{\ssrm LU}}$, Eq.~(\ref{eq8}) is
 in fact a differential equation of redshift $z$ with respect
 to $\tau$.~Solving this differential equation with the initial
 condition $\tau=1$ (or equivalently $t=t_0$) at $z=0$, we have
 \be{eq9}
 -\ln (1+z)=\xi_0\ln\tau+\frac{\xi_0}{2}\left(1-\tau\right)
 \left[\left(1-\tau\right)\eta_{\ssrm LU}+2-\frac{2}{\xi_0}\,
 p_{\rm age,\,{\ssrm LU}}\right]\,.
 \ee
 One can numerically obtain the positive $\tau(z)$ as a function of
 redshift $z$ from Eq.~(\ref{eq9}), and then $E_{\ssrm LU}(z)$ as well
 as $H_{\ssrm LU}(z)=H_{0,\,{\ssrm LU}}E_{\ssrm LU}(z)$ are on hand by
 using Eq.~(\ref{eq8}).

The next step is to match these two segments at the critical redshift
 $z=z_{\rm cr}$, where the physical quantities $t$ (or equivalently
 $\tau=t/t_0$) and $H$ should be continuous, respectively (and hence $Ht$ is
 also continuous naturally, n.b.~Eq.~(\ref{eq5})).~Note that $E$ is
 not continuous at $z=z_{\rm cr}$ if $H_{0,\,{\ssrm EU}}\not=H_{0,\,
 {\ssrm LU}}$.~At the critical redshift $z_{\rm cr}$, we can numerically
 obtain the corresponding $\tau_{\rm cr}=\tau(z=z_{\rm cr})$
 from Eq.~(\ref{eq7}), namely
 \be{eq10}
 -\ln (1+z_{\rm cr})=\alpha_0\ln\tau_{\rm cr}+\frac{\alpha_0}{2}\left(1-
 \tau_{\rm cr}\right)\left[\left(1-\tau_{\rm cr}\right)\eta_{\ssrm EU}+2
 -\frac{2}{\alpha_0}\,p_{\rm age,\,{\ssrm EU}}\right]\,.
 \ee
 Noting the match condition for $t$ (or equivalently $\tau=t/t_0$) and
 substituting this known $\tau_{\rm cr}$ into Eq.~(\ref{eq9}), we have
 \be{eq11}
 \xi_0=\left[\left(1-\tau_{\rm cr}\right) p_{\rm age,\,{\ssrm LU}}-
 \ln\left(1+z_{\rm cr}\right)\,\right]\cdot\left[\left(1-\tau_{\rm cr}
 \right)+\frac{\eta_{\ssrm LU}}{2}\left(1-
 \tau_{\rm cr}\right)^2+\ln\tau_{\rm cr}\right]^{-1}\,.
 \ee
 So, $\xi_0$ is not an independent model parameter.~The second match
 condition is for $H$, namely $H_{\ssrm EU}(z_{\rm cr})
 =H_{\ssrm LU}(z_{\rm cr})=H_{0,\,{\ssrm EU}}E_{\ssrm EU}(z_{\rm cr})=
 H_{0,\,{\ssrm LU}}E_{\ssrm LU}(z_{\rm cr})$. Using Eqs.~(\ref{eq6}) and
 (\ref{eq8}), we find
 \be{eq12}
 \eta_{\ssrm LU}=\frac{1}{\tau_{\rm cr}}\left(1-\frac{\gamma}{\xi_0}
 \right)\,,\quad {\rm where}\quad\gamma\equiv\frac{\left(
 \,p_{\rm age,\,{\ssrm EU}}-p_{\rm age,\,{\ssrm LU}}\right)\tau_{\rm cr}}{1-
 \tau_{\rm cr}}+\alpha_0\left(1-\eta_{\ssrm EU}\,\tau_{\rm cr}\right)\,.
 \ee
 Substituting this $\eta_{\ssrm LU}$ into the
 right hand side of Eq.~(\ref{eq11}), we finally obtain
 \be{eq13}
 \xi_0=\left[\left(1-\tau_{\rm cr}\right)p_{\rm age,\,{\ssrm
 LU}}+\frac{\gamma\left(1-\tau_{\rm cr}\right)^2}{2\,\tau_{\rm cr}}-
 \ln\left(1+z_{\rm cr}\right)\right]\cdot\left(\ln\tau_{\rm cr}
 +\frac{1-\tau_{\rm cr}^2}{2\,\tau_{\rm cr}}\right)^{-1}\,.
 \ee
 Substituting this $\xi_0$ into Eq.~(\ref{eq12}), the final formula for
 $\eta_{\ssrm LU}$ is ready.~Both $\xi_0$ and $\eta_{\ssrm LU}$ are not
 independent model parameters.~In fact, $p_{\rm age,\,{\ssrm LU}}$ is
 also not an independent model parameter, since
 \be{eq14}
 p_{\rm age,\,{\ssrm LU}}=\frac{H_{0,\,{\ssrm LU}}}{H_{0,\,{\ssrm EU}}}
 \cdot p_{\rm age,\,{\ssrm EU}}\,.
 \ee
 So, only $\alpha_0$, $\eta_{\ssrm EU}$, $p_{\rm age,\,{\ssrm EU}}$,
 $H_{0,\,{\ssrm EU}}$, $H_{0,\,{\ssrm LU}}$ and $z_{\rm cr}$ are independent
 model parameters.

In the present work, we consider two types of TGPAge.~If all the independent
 model parameters are free, we label it as TGPAge-Free.~If
 $H_{0,\,{\ssrm EU}}$ and $H_{0,\,{\ssrm LU}}$ are fixed, we label it as
 TGPAge-Fix.~As mentioned above, $H_{0,\,{\ssrm EU}}$ and
 $H_{0,\,{\ssrm LU}}$ correspond to the Hubble constant inferred from
 CMB and the Hubble constant measured by SH0ES with nearby
 Cepheids/SNIa, respectively.~Thus, we fix $H_{0,\,{\ssrm EU}}=67.36\;
 {\rm km/s/Mpc}$~\cite{Planck:2018vyg} and $H_{0,\,{\ssrm LU}}=73.04
 \;{\rm km/s/Mpc}$~\cite{Riess:2021jrx} in TGPAge-Fix, respectively.


 \begin{table}[tb]
 \renewcommand{\arraystretch}{1.6}
 \begin{center}
 \vspace{-5mm}  
 \begin{tabular}{c} \hline\hline
 $\left|\,\Delta {\rm AIC}\,\right|$\\ \hline
 \renewcommand{\arraystretch}{1.6}
 \quad\ Level of empirical support for the model with the smaller AIC\quad\ \ \\ \hline
 \renewcommand{\arraystretch}{1.0}
 \begin{tabular}{ccc}\\[-3.6mm]
 $0-2$\hspace{10mm} &$4-7$\hspace{10mm} &$>10$\\[-0.3mm]
 Weak\hspace{10mm} &Mild\hspace{10mm} &Strong\\[1mm]
 \end{tabular}
 \\ \hline\hline
 $\left|\,\Delta {\rm BIC}\,\right|$\\ \hline
 \renewcommand{\arraystretch}{1.6}
 Evidence against the model with the larger BIC\\ \hline
 \renewcommand{\arraystretch}{1.0}
 \begin{tabular}{cccc}\\[-3.6mm]
 $0-2$\hspace{10mm} &$2-6$\hspace{10mm} &$6-10$\hspace{10mm} &$>10$\\[-0.3mm]
 Weak\hspace{10mm} &Positive\hspace{10mm} &Strong\hspace{10mm} &Very strong\\[1mm]
 \end{tabular}
 \\ \hline\hline
 $\left|\,\ln {\cal B}\,\right|$\\ \hline
 \renewcommand{\arraystretch}{1.6}
 Evidence against the model with the smaller $\cal Z$\\ \hline
 \renewcommand{\arraystretch}{1.0}
 \begin{tabular}{cccc}\\[-3.6mm]
 $0-1$\hspace{10mm} & $1-2.5$\hspace{10mm} & $2.5-5$\hspace{10mm} & $>5$\\[-0.3mm]
 Inconclusive\hspace{10mm} &Weak\hspace{10mm} & Moderate\hspace{10mm} & Strong\\[1mm]
 \end{tabular}
 \\ \hline\hline
 \end{tabular}
 \end{center}
 \vspace{-1mm}  
 \caption{\label{tab2} The empirical strength of $\Delta$AIC,
 $\Delta$BIC and $\ln {\cal B}$
 (see e.g.~\cite{Jia:2025prq,Jia:2025kvp}).}
 \end{table}



\section{Observational data}\label{sec3}

Here, we briefly introduce the observational data used in the present
 work.~The Pantheon+ SNIa sample~\cite{Brout:2022vxf,Scolnic:2021amr,
 PantheonPlusSH0ES} consists of 1701 light curves of 1550 SNIa
 at redshits $0.00122\leq z\leq 2.26137$ (in this work we use the Hubble
 diagram redshifts $z_{\rm HD}$ given by column 3 of the Pantheon+ data
 table with CMB and peculiar velocity corrections).~Note that 77 Cepheid
 calibrated host-galaxy distance moduli are also provided by SH0ES at
 redshifts $0.00122\leq z\leq 0.01682$.~Following~\cite{Brout:2022vxf},
 the $\chi^2$ from the Pantheon+ SNIa sample is given by
 \be{eq15}
 \chi^2_{\rm SN}=\Delta{\boldsymbol{\mu}}^{\,T}
 \cdot\boldsymbol{C}_{\rm SN,\,stat+sys}^{\,-1}
 \cdot\Delta{\boldsymbol{\mu}}\,,
 \ee
 where $\boldsymbol{C}_{\rm SN,\,stat+sys}$ is the covariance matrix
 including statistical and systematic uncertainties, $\Delta
 {\boldsymbol{\mu}}$ is the vector of 1701 SNIa distance
 modulus residuals computed as
 \be{eq16}
 \Delta\mu_i=
 \begin{cases}\\[-5.8mm]
 \;\mu_i-\mu_i^{\rm Ceph} & i\in {\rm Cepheid\ hosts}\,,\\[1.8mm]
 \;\mu_i-\mu_{\rm model}(z_i) \quad & {\rm otherwise}\,,\\[-1.411mm]
 \end{cases}
 \ee
 in which $\mu_i=m_{B,\,i}-M$ is the distance modulus of the $i$-th
 SNIa, $M$ is the absolute magnitude, $\mu_i^{\rm Ceph}$ is the
 Cepheid calibrated host-galaxy distance modulus provided by SH0ES,
 $m_{B,\,i}$ is the corrected/standardized apparent magnitude, and the
 model distance modulus $\mu_{\rm model}(z_i)$ is given by
 \be{eq17}
 \mu_{\rm model}(z_i)=
 5\log\left(d_L(z_i)/{\rm Mpc}\right)+25\,,
 \ee
 while ``\;$\log$\,'' gives the logarithm to base $10$, and $d_L(z)=
 \left(1+z\right)r(z)$ is the luminosity distance predicted
 by the model.~The theoretical comoving distance $r(z)$
 is given by~\cite{Weinberg:2008zzc}
 \be{eq18}
 r(z)=c\int_0^z\frac{d\tilde{z}}{H(\tilde{z})}=
 c\int_0^z\frac{d\tilde{z}}{H_0 E(\tilde{z})}\,.
 \ee
 where $c$ is the speed of light.~Noting that $H(z)=H_{\ssrm LU}(z)$ and
 $H_{\ssrm EU}(z)$ for $z\leq z_{\rm cr}$ and $z\geq z_{\rm cr}$ in
 TGPAge respectively, its comoving distance $r(z)$ reads


 \begin{table}[tb]
 \renewcommand{\arraystretch}{1.6}
 \begin{center}
 \vspace{-3mm}  
 \hspace{-3mm}  
 \begin{tabular}{lc} \hline\hline
 Model & Uniform priors \\ \hline
 $\Lambda$CDM & $M\in [\,-21,\, -18\,]$,\quad $H_0\in [\,50,\, 90\,]$\quad (and\quad $\Omega_{m0}\in [\,0.1,\, 0.5\,]$\quad if it is free)\\
 SGPAge & $M\in [\,-21,\, -18\,]$,\quad $H_0\in [\,50,\, 90\,]$,\quad $\alpha_0\in [\,0.1,\, 1\,]$,\quad $\eta\in [\,-1,\, 1\,]$,\quad $p_{\rm age}\in [\,0.1, 2\,]$\\
 TGPAge-Fix\hspace{5mm} & $M\in [\,-21,\, -18\,]$,\quad $\alpha_0\in [\,0.1,\, 1\,]$,\quad $\eta_{\ssrm EU}\in [\,-1,\, 1\,]$,\quad
 $p_{\rm age,\,\ssrm EU}\in [\,0.1,\, 2\,]$,\quad $z_{\rm cr}\in [\,0.001,\, 5\,]$ \\
 TGPAge-Free & $M\in [\,-21,\, -18\,]$,\quad $H_{0,\,\ssrm LU}\in [\,50,\, 90\,]$,\quad $H_{0,\,\ssrm EU}\in [\,20,\, 80\,]$,\quad $\alpha_0\in [\,0.1,\, 1\,]$\\[-1mm]
  & $\eta_{\ssrm EU}\in [\,-5,\, 5\,]$,\quad $p_{\rm age,\,\ssrm EU}\in [\,0.1,\, 2\,]$,\quad $z_{\rm cr}\in [\,0.001,\, 10\,]$ \\[0.4mm]
 \hline\hline
 \end{tabular}
 \end{center}
 \vspace{-1mm}  
 \caption{\label{tab3} The uniform priors for all the models considered
 in this work.~See Sec.~\ref{sec4} for details.}
 \end{table}


\vspace{-4.5mm}  

 \bea
 r(z)=c\int_0^z\frac{d\tilde{z}}{H_{0,\,\ssrm LU}E_{\ssrm LU}(\tilde{z})}
 \quad {\rm if}\quad z\leq z_{\rm cr}\,,\label{eq19}\\[1.5mm]
 r(z)=c\int_0^{z_{\rm cr}}\frac{d\tilde{z}}{H_{0,\,\ssrm LU}
 E_{\ssrm LU}(\tilde{z})}+c\int_{z_{\rm cr}}^z\frac{d\tilde{z}}
 {H_{0,\,\ssrm EU}E_{\ssrm EU}(\tilde{z})}
 \quad {\rm if}\quad z>z_{\rm cr}\,.\label{eq20}
 \eea
 It is worth noting that a new model parameter $M$ is introduced to
 use SNIa data.~As is well known, in most SNIa samples, the absolute
 magnitude $M$ and the Hubble constant $H_0$ are heavily degenerated
 through a combination ${\cal M}=M+5\log\left(c/H_0/{\rm Mpc}\right)
 +25$.~But in the Pantheon+ SNIa sample, the degeneracy between $M$ and
 $H_0$ can be broken to some extent, by using the 77 Cepheid calibrated
 host-galaxy distance moduli provided by SH0ES, which can be used to
 constrain the absolute magnitude $M$ alone (n.b.~the first
 row of Eq.~(\ref{eq16}), regardless of $H_0$ appeared
 in $d_L$ of the theoretical model).

Since the Hubble tension is mainly between CMB and SNIa/Cepheids, it is
 necessary to also consider the observational data of CMB.~It consumes a
 large amount of time and power to use the full CMB data.~As an alternative,
 the distance priors derived from the full CMB data have been
 extensively used the literature, which contain the main information of
 CMB.~Here, we consider the distance priors in~\cite{Chen:2018dbv} derived
 from the CMB observation of Planck 2018.~In this
 case, the $\chi^2$ from CMB is given by~\cite{Chen:2018dbv}
 \be{eq21}
 \chi^2_{\rm CMB}=\Delta\boldsymbol{d}^{\,T}
 \cdot\boldsymbol{C}_{d}^{\,-1}\cdot\Delta\boldsymbol{d}\,,
 \ee
 where $\Delta\boldsymbol{d}$ is the vector of distance prior
 residuals, and $\boldsymbol{C}_{d}^{\,-1}$ is the inverse covariance
 matrix, namely
 \be{eq22}
 \Delta\boldsymbol{d}=\left(\begin{array}{c}
                              {\cal R}_{\rm model}(z_\ast)-1.750235\\
                              \ell_{A,\,{\rm model}}(z_\ast)-301.4707
                            \end{array}\right)\,,\hspace{10mm}
 \boldsymbol{C}_{d}^{\,-1}=\left(\begin{array}{lr}
                                   94392.3971 & \ -1360.4913\\
                                   -1360.4913 & 161.4349
                            \end{array}\right)\,,
 \ee
 in which $z_\ast=1089.92$ from the Planck 2018
 result~\cite{Planck:2018vyg}.~The shift parameter $\cal R$ and the
 acoustic scale $\ell_A$ are
 given by (e.g.~\cite{Wang:2007mza,Wang:2006ts} and~\cite{Chen:2018dbv})
 \be{eq23}
 {\cal R}=\left(\Omega_{m0}H_0^2\right)^{1/2}r(z_\ast)/c\,,
 \hspace{10mm}\ell_A=\pi r(z_\ast)/r_s(z_\ast)\,,
 \ee
 where the comoving distance $r(z)$ is given by Eqs.~(\ref{eq18}) or
 (\ref{eq20}) (note that $z_\ast\gg z_{\rm cr}$ in TGPAge, and $H_0$
 in $\cal R$ should be regarded as $H_{0,\,\ssrm EU}$ in
 TGPAge).~The comoving sound horizon $r_s(z)$ is given
 by (e.g.~\cite{Wang:2007mza,Wright:2007vr} and \cite{Chen:2018dbv})
 \vspace{-2mm}  
 \be{eq24}
 r_s(z)=c\int_0^{1/(1+z)}\frac{c_s\,da}{a^2 H(a)}\,,
 \ee
 where the sound speed $c_s$ reads~\cite{Hu:1995en} (see
 also e.g.~\cite{Chen:2018dbv,Wang:2007mza})
 \be{eq25}
 c_s(a)=1/\sqrt{3\left(1+R(a)\right)}\,,\quad {\rm and}\quad
 R(a)=\frac{3}{4}\frac{\rho_b}{\rho_\gamma}=
 \frac{3(\Omega_b h^2)\,a}{4(\Omega_\gamma h^2)}=
 31500\left(T_{\rm CMB}/2.7\,{\rm K}\right)^{-4}(\Omega_b h^2)\,a\,,
 \ee
 in which $a=(1+z)^{-1}$ is the scale factor,
 $T_{\rm CMB}=2.7255\,{\rm K}$~\cite{Fixsen:2009ug}, and
 $\Omega_b h^2=0.02237$ from the Planck 2018
 result~\cite{Planck:2018vyg}, while $h=H_0/(100\;{\rm km/s/Mpc})$.~In
 TGPAge, $H=H_{\ssrm EU}=H_{0,\,\ssrm EU}E_{\ssrm EU}$ in
 Eq~(\ref{eq24}) since $z_\ast\gg z_{\rm cr}$.~Note that the present
 fractional matter density $\Omega_{m0}$ does not explicitly appear in
 all PAge parameterizations (n.b.~Sec.~\ref{sec2}), and we should find
 a way out.~In fact, for the given $z_\ast=1089.92$ from the Planck 2018
 result~\cite{Planck:2018vyg}, noting that~\cite{Hu:1995en,Chen:2018dbv}
 \bea
 z_\ast=1048\left[\,1+0.00124\,(\Omega_b h^2)^{-0.738}\,\right]\cdot
 \left[\,1+g_1\,(\Omega_{m0} h^2)^{\,g_2}\right]\,,\label{eq26}\\[1mm]
 g_1=\frac{0.0738\,(\Omega_b h^2)^{-0.238}}{1+39.5\,(\Omega_b
 h^2)^{\,0.763}}\,,\hspace{12mm} g_2=\frac{0.560}{1+21.1\,(\Omega_b
 h^2)^{\,1.81}}\,,\label{eq27}
 \eea
 we can infer $\Omega_{m0} h^2$ by using Eqs.~(\ref{eq26}) and
 (\ref{eq27}) with $\Omega_b h^2=0.02237$ from the Planck 2018
 result~\cite{Planck:2018vyg}, and then $\sqrt{\Omega_{m0}
 H_0^2\,}=\sqrt{\Omega_{m0} h^2\,}\times 100\;{\rm km/s/Mpc}$ in $\cal R$ is
 ready (note that $h\,$ should be regarded as~$h_{\ssrm EU}=H_{0,\,\ssrm
 EU}/(100\;{\rm km/s/Mpc})$ in TGPAge since $z_\ast\gg z_{\rm cr}$).

Here, we also consider the data of baryon acoustic oscillations
 (BAO).~Recently, DESI BAO DR2 data have been released~\cite{DESI:2025zgx,
 DESIBAO}, which provide the $D_{\rm M}/r_d$, $D_{\rm H}/r_d$
 and $D_{\rm V}/r_d$ data for some tracers, where $D_{\rm M}(z)=r(z)$,
 $D_{\rm H}(z)=c/H(z)$, $D_{\rm V}(z)=(zD_{\rm M}^2(z)D_{\rm H}(z))^{1/3}$,
 and $r_d$ is the sound horizon at drag epoch. One might directly use these
 BAO data, but a new free model parameter $r_d$ will be introduced, which is
 heavily degenerated with the Hubble constant $H_0$ in using BAO data
 through a combination $H_0 r_d$ as well known.~If we use the SNIa and BAO
 data jointly, the situation becomes fairly worse, since $H_0$ is also
 degenerated with the absolute magnitude $M$ of SNIa data as
 mentioned above.~Thus, it is much better to consider the BAO
 data without $r_d\,$.~Following e.g.~\cite{DESI:2024mwx,Wang:2024pui,
 Gao:2025ozb,Gong:2025hoy}, we instead use the Alcock-Paczynski (AP)
 factor $F_{\rm AP}$ in this work, namely


 \begin{center}
 \begin{figure}[tb]
 \centering
 \vspace{-7mm}  
 \includegraphics[width=0.45\textwidth]{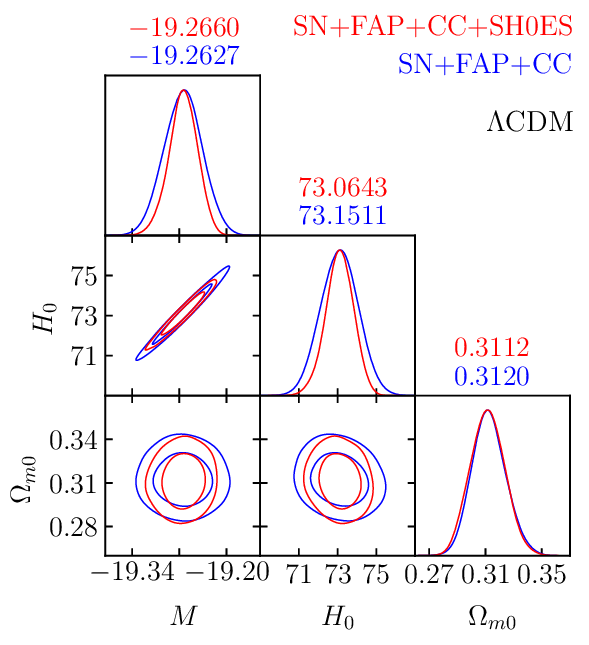}
 \vspace{-2mm}  
 \caption{\label{fig2} The $1\sigma$ and $2\sigma$ contours for all the
 free parameters of the $\Lambda$CDM model from the observational data
 SN +FAP+CC (blue) and SN+FAP+CC+SH0ES (red).~The marginalized
 probability distributions and the best-fit values are also given at
 the tops of all columns for the corresponding parameters.~The Hubble
 constant $H_0$ is in units of $\rm km/s/Mpc$.~See Sec.~\ref{sec4} for
 details.}
 \end{figure}
 \end{center}


\vspace{-14mm} 

 \be{eq28}
 F_{\rm AP}\equiv
 \frac{D_{\rm M}}{D_{\rm H}}=\frac{D_{\rm M}/r_d}{D_{\rm H}/r_d}\,,
 \ee
 in which $r_d$ has been canceled.~Using the numerical data table and
 the covariance matrix of $D_{\rm M}/r_d$ and $D_{\rm H}/r_d$ for DESI
 BAO DR2 given in~\cite{DESIBAO}, we obtain 6 data points of $F_{\rm
 AP}$ and their uncertainties, through the propagation of uncertainties with
 the Jacobian matrix.~We present them in Table~\ref{tab1}.~The $\chi^2$
 from $F_{\rm AP}$ is given by


 \begin{table}[tb]
 \renewcommand{\arraystretch}{1.6}
 \begin{center}
 \vspace{-3mm}  
 \hspace{-3mm}  
 \begin{tabular}{lccc} \hline\hline
 Data & $M$ & $H_0$ & $\Omega_{m0}$ \\ \hline
 SN+FAP+CC (SFC) & $-19.2643^{+0.0278}_{-0.0287}$ & $73.0973^{+0.9578}_{-0.9779}$ & $0.3125^{+0.0111}_{-0.0128}$ \\
 SN+FAP+CC+SH0ES (SFCS) & $-19.2643^{+0.0222}_{-0.0198}$ & $73.1038^{+0.7146}_{-0.6985}$ & $0.3115^{+0.0124}_{-0.0123}$ \\
 SN+CMB (SC) & $-19.3477^{+0.0045}_{-0.0046}$ & $70.7743^{+0.0952}_{-0.0960}$ & \\
 SN+CMB+FAP (SCF) & $-19.3479^{+0.0043}_{-0.0045}$ & $70.7670^{+0.0910}_{-0.0937}$ & \\
 SN+CMB+FAP+CC (SCFC) & $-19.3479^{+0.0044}_{-0.0045}$ & $70.7678^{+0.0944}_{-0.0899}$ & \\
 SN+CMB+FAP+CC+SH0ES (SCFCS)\hspace{5mm} & $-19.3473^{+0.0047}_{-0.0042}$ & \hspace{5mm}$70.7899^{+0.0968}_{-0.1023}$\hspace{5mm} & \\[0.2mm]
 \hline\hline
 \end{tabular}
 \end{center}
 \vspace{-1mm}  
 \caption{\label{tab4} The means and $1\sigma$ intervals for all the
 free parameters of the $\Lambda$CDM model by using various
 observational data.~See Sec.~\ref{sec4} for details.}
 \end{table}


\vspace{-5mm}   

 \be{eq29}
 \chi^2_{\rm FAP}=\sum_i\left(\frac{F_{\rm AP}^{\rm obs}(z_i)-
 F_{{\rm AP}}^{\rm model}(z_i)}{\sigma_{F_{\rm AP},\,i}}\right)^2\,.
 \ee

Since all the PAge parameterizations directly parameterize the Hubble
 parameter $H(\tau)$ and then give $H(z)$ as a function of redshift $z$
 (n.b.~Sec.~\ref{sec2}), it is natural and suitable to use the
 observational $H(z)$ data to constrain them.~As is well known, the
 cosmic chronometers (CC), namely the measurements of $H(z)$ from the
 relative ages of massive early-type passively evolving galaxies, were
 firstly proposed in~\cite{Jimenez:2001gg,Jimenez:2003iv,Simon:2004tf}
 (see also e.g.~\cite{Samushia:2006fx,Yi:2006bw,Wei:2006ut,Wei:2007ws,
 Yin:2018mvu}), and have been developed for 25 years.~In this work, we
 use the 32 CC data sample given by Table~1 of~\cite{Moresco:2022phi} in
 the redshift range of $0.07\leq z\leq 1.965$, and the covariance matrix
 $\boldsymbol{C}_{\rm CC,\,stat+sys}$ including statistical and
 systematic uncertainties can be found from~\cite{CCcov}
 (see~\cite{Moresco:2020fbm} and Sec.~3.1 of~\cite{Moresco:2022phi} for
 details).~The $\chi^2$ from CC is given by
 \be{eq30}
 \chi^2_{\rm CC}=\Delta{\boldsymbol{H}}^{\,T}
 \cdot\boldsymbol{C}_{\rm CC,\,stat+sys}^{\,-1}
 \cdot\Delta{\boldsymbol{H}}\,,
 \ee
 where $\Delta{\boldsymbol{H}}$ is the vector of $H$ residuals
 computed as $\Delta H_i=H_{\rm obs}(z_i)-H_{\rm model}(z_i)$.

Finally, the local and direct measurement $H_0=73.04\pm
 1.04\;{\rm km/s/Mpc}$ of SH0ES~\cite{Riess:2021jrx} by using nearby
 Cepheid/SNIa can also be considered as an independent data point.~The
 $\chi^2$ from SH0ES is given by
 \be{eq31}
 \chi^2_{\rm SH0ES}=\left(\frac{H_0^{\rm model}-73.04}{1.04}\right)^2\,,
 \ee
 where $H_0^{\rm model}$ is in units of $\rm km/s/Mpc$.~Note that in TGPAge,
 this $H_0$ should be regarded as $H_{0,\,\ssrm LU}$ since $z=0<z_{\rm
 cr}$.~We consider that it might be slightly controversial to use this as an
 independent data point, and hence the results from this data point are
 for reference only.

For model comparison, the Akaike and the Bayesian information criteria
 AIC, BIC and the Bayesian evidence have been extensively used in the
 literature.~AIC~\cite{Akaike:1974} and BIC~\cite{Schwarz:1978} are
 defined by
 \be{eq32}
 {\rm AIC}=-2\ln {\cal L}_{\rm max}+2\kappa\,,\hspace{10mm}
 {\rm BIC}=-2\ln {\cal L}_{\rm max}+\kappa\ln N\,,
 \ee
 where ${\cal L}_{\rm max}=\exp\,(-\chi^2_{\rm min}/2)$ is the
 maximum likelihood, $N$ and $\kappa$ are the numbers of data points and
 free model parameters, respectively.~We compare two models ${\cal M}_1$ and
 ${\cal M}_2$ by computing $\Delta {\rm AIC}_{12}={\rm AIC}_1-{\rm AIC}_2$
 and $\Delta {\rm BIC}_{12}={\rm BIC}_1-{\rm BIC}_2$.~A negative (positive)
 $\Delta {\rm AIC}_{12}$ or $\Delta {\rm BIC}_{12}$ means a preference
 for ${\cal M}_1$ (${\cal M}_2$).~The Bayesian evidence is defined by
 (e.g.~\cite{Kass:1995loi,Kilbinger:2009by,Weinberg:2009rd,
 Trotta:2008qt,Mukherjee:2017oom})
 \be{eq33}
 {\cal Z}=\int {\cal L}(\boldsymbol{\psi})\,P(\boldsymbol{\psi})\,d
 \boldsymbol{\psi}\,,
 \ee
 where $\cal L$ is the likelihood function, $P$ is the prior
 distribution, and $\boldsymbol{\psi}$ denotes the model parameters.~We
 compare two models ${\cal M}_1$ and ${\cal M}_2$ by computing the Bayes
 factor ${\cal B}_{12}={\cal Z}_1/{\cal Z}_2$ or equivalently $\ln {\cal
 B}_{12}=\ln {\cal Z}_1 -\ln {\cal Z}_2$.~A positive (negative)
 $\ln {\cal B}_{12}$ means a preference for ${\cal M}_1$ (${\cal M}_2$).~The
 strengthes of evidences are indicated by the empirical ranges
 of $\left|\hspace{0.2mm}\Delta {\rm AIC}\hspace{0.3mm}\right|$, $\left|
 \hspace{0.2mm}\Delta {\rm BIC}\hspace{0.3mm}\right|$,
 $\left|\,\ln {\cal B}\,\right|$ summarized in Table~\ref{tab2}
 (see e.g.~\cite{Jia:2025prq,Jia:2025kvp}).


 \begin{center}
 \begin{figure}[tb]
 \centering
 \vspace{-7mm}  
 \includegraphics[width=0.6\textwidth]{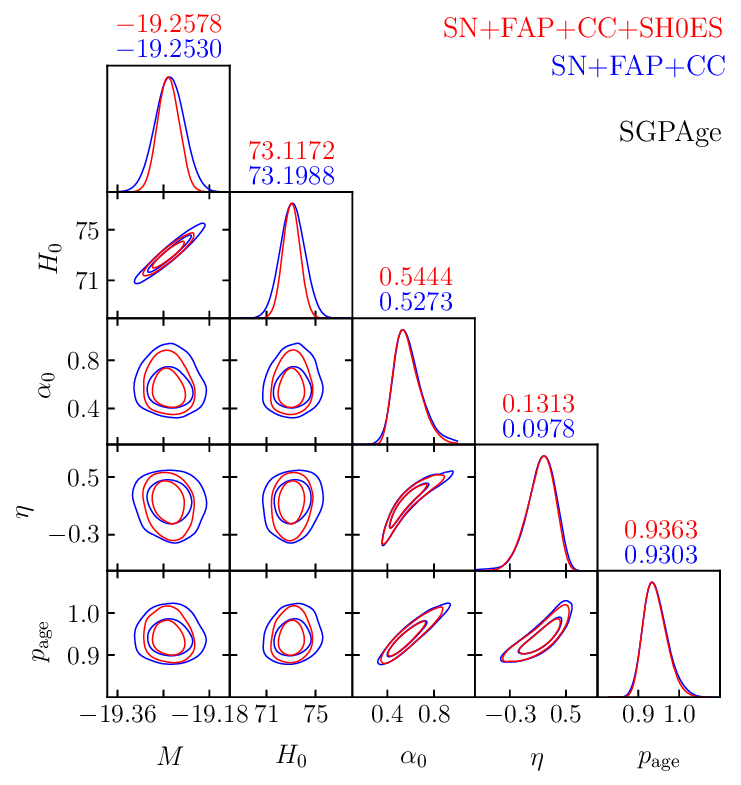}
 \vspace{-2mm}  
 \caption{\label{fig3} The same as in Fig.~\ref{fig2}, but for
 SGPAge.~See Sec.~\ref{sec4} for details.}
 \end{figure}
 \end{center}


\vspace{-8mm} 

We minimize $\chi^2$ by using the Markov Chain Monte Carlo
 (MCMC) Python package Cobaya~\cite{Torrado:2020dgo,Cobaya}
 with GetDist~\cite{Lewis:2019xzd,GetDist}, and then obtain the
 constraints on the model parameters.~We compute the Bayesian
 evidence by using MCEvidence with
 the MCMC chains~\cite{Heavens:2017afc,MCEvidence,MCEvimod}.


 \begin{sidewaystable}[tbp] 
 \renewcommand{\arraystretch}{2}
 \begin{center}
 \vspace{-2mm}  
 \hspace{1mm}   
 \begin{tabular}{llccccccc}\hline\hline
 Data & Model & $M$ & $H_{0,\,\ssrm LU}$ & $H_{0,\,\ssrm EU}$ & $\alpha_0$ & $\eta_{\ssrm EU}$ & $p_{\rm age,\,\ssrm EU}$ & $z_{\rm cr}$ \\[1mm] \hline

 SFC & SGPAge & $-19.2578^{+0.0284}_{-0.0282}$ & $73.1132^{+0.9789}_{-0.9701}$ & & $0.5811^{+0.0850}_{-0.1395}$ & $0.1428^{+0.2381}_{-0.1639}$ & $0.9430^{+0.0237}_{-0.0338}$ & \\
 SFCS & SGPAge & $-19.2601^{+0.0212}_{-0.0206}$ & $73.0358^{+0.6820}_{-0.6827}$ & & $0.5733^{+0.0792}_{-0.1329}$ & $0.1383^{+0.2272}_{-0.1615}$ & $0.9415^{+0.0227}_{-0.0318}$ & \\
 SC & SGPAge & $-19.2490^{+0.0292}_{-0.0287}$ & $73.2612^{+0.9562}_{-1.0188}$ & & $0.6208^{+0.0010}_{-0.0010}$ & $0.0190^{+0.0599}_{-0.0588}$ & $1.0141^{+0.0091}_{-0.0091}$ & \\
 SCF & SGPAge & $-19.3044^{+0.0263}_{-0.0286}$ & $71.3517^{+0.8379}_{-0.9653}$ & & $0.6220^{+0.0009}_{-0.0009}$ & $0.0443^{+0.0555}_{-0.0568}$ & $1.0002^{+0.0080}_{-0.0089}$ & \\
 SCFC & SGPAge & $-19.3101^{+0.0270}_{-0.0268}$ & $71.1550^{+0.8564}_{-0.8665}$ & & $0.6221^{+0.0009}_{-0.0009}$ & $0.0493^{+0.0576}_{-0.0581}$ & $0.9990^{+0.0082}_{-0.0082}$ & \\
 SCFCS & SGPAge & $-19.2867^{+0.0206}_{-0.0210}$ & $71.9484^{+0.6705}_{-0.6854}$ & & $0.6216^{+0.0008}_{-0.0008}$ & $0.0383^{+0.0565}_{-0.0547}$ & $1.0046^{+0.0074}_{-0.0073}$ & \\[0.6mm] \hline

 SFC & TGPAge-Fix & $-19.2599^{+0.0081}_{-0.0081}$ & & & $0.5441^{+0.0676}_{-0.1291}$ & $-0.3176^{+0.3179}_{-0.2316}$ & $0.8668^{+0.0192}_{-0.0293}$ & $3.1825^{+1.3832}_{-0.9827}$ \\
 SFCS & TGPAge-Fix & $-19.2598^{+0.0082}_{-0.0082}$ & & & $0.5451^{+0.0684}_{-0.1319}$ & $-0.3155^{+0.3217}_{-0.2298}$ & $0.8668^{+0.0186}_{-0.0303}$ & $3.2264^{+1.7736}_{-0.6164}$ \\
 SC & TGPAge-Fix & $-19.2579^{+0.0082}_{-0.0082}$ & & & $0.6186^{+0.0012}_{-0.0010}$ & $-0.3585^{+0.0836}_{-0.0728}$ & $0.9260^{+0.0072}_{-0.0065}$ & $2.7502^{+1.0123}_{-1.6443}$ \\
 SCF & TGPAge-Fix & $-19.2597^{+0.0080}_{-0.0078}$ & & & $0.6183^{+0.0009}_{-0.0009}$ & $-0.3790^{+0.0624}_{-0.0616}$ & $0.9243^{+0.0056}_{-0.0057}$ & $1.9154^{+0.3463}_{-0.6602}$ \\
 SCFC & TGPAge-Fix & $-19.2619^{+0.0081}_{-0.0075}$ & & & $0.6185^{+0.0008}_{-0.0008}$ & $-0.3685^{+0.0597}_{-0.0589}$ & $0.9252^{+0.0054}_{-0.0054}$ & $1.8308^{+0.3367}_{-0.6379}$ \\
 SCFCS & TGPAge-Fix & $-19.2613^{+0.0077}_{-0.0079}$ & & & $0.6184^{+0.0008}_{-0.0008}$ & $-0.3711^{+0.0589}_{-0.0580}$ & $0.9250^{+0.0053}_{-0.0054}$ & $1.8301^{+0.3213}_{-0.6223}$ \\[0.6mm] \hline

 SFC & TGPAge-Free & $-19.2597^{+0.0274}_{-0.0271}$ & $73.0407^{+0.9465}_{-0.9373}$ & $56.8899^{+23.1101}_{-7.6616}$
 & $0.5479^{+0.0882}_{-0.1299}$ & $-0.9456^{+1.5136}_{-0.5988}$ & $0.7331^{+0.2817}_{-0.1366}$ & $5.6596^{+4.3404}_{-1.5109}$ \\
 SFCS & TGPAge-Free & $-19.2598^{+0.0210}_{-0.0214}$ & $73.0357^{+0.7095}_{-0.7067}$ & $56.5441^{+23.4559}_{-7.7372}$
 & $0.5483^{+0.0873}_{-0.1338}$ & $-0.9802^{+1.5394}_{-0.6087}$ & $0.7294^{+0.2879}_{-0.1383}$ & $5.6893^{+2.9413}_{-2.8543}$ \\
 SC & TGPAge-Free & $-19.2486^{+0.0286}_{-0.0289}$ & $73.3809^{+0.9823}_{-1.0069}$ & $54.5117^{+22.8551}_{-9.9095}$
 & $0.6167^{+0.0042}_{-0.0029}$ & $-1.0220^{+1.0739}_{-0.7023}$ & $0.7474^{+0.2855}_{-0.1675}$ & $5.5240^{+2.9087}_{-2.7983}$ \\
 SCF & TGPAge-Free & $-19.2561^{+0.0291}_{-0.0295}$ & $73.2878^{+1.0147}_{-1.0261}$ & $39.3967^{+11.7970}_{-12.2583}$
 & $0.6116^{+0.0024}_{-0.0026}$ & $-1.9598^{+0.5409}_{-0.6535}$ & $0.5307^{+0.1602}_{-0.1665}$ & $3.7030^{+0.6806}_{-1.1745}$ \\
 SCFC & TGPAge-Free & $-19.2675^{+0.0281}_{-0.0284}$ & $72.9330^{+0.9935}_{-1.0021}$ & $39.6590^{+10.9561}_{-13.2241}$
 & $0.6123^{+0.0023}_{-0.0026}$ & $-1.8996^{+0.5391}_{-0.6914}$ & $0.5372^{+0.1458}_{-0.1834}$ & $3.7561^{+0.7662}_{-1.1278}$ \\
 SCFCS~~~~ & TGPAge-Free~~~~ & $-19.2664^{+0.0213}_{-0.0213}$ & ~~~~$72.9722^{+0.7127}_{-0.7264}$~~~~ & $39.2265^{+10.5418}_{-13.6647}$
 & ~~~~$0.6123^{+0.0021}_{-0.0023}$~~~~ & $-1.9228^{+0.5089}_{-0.7079}$ & ~~~~$0.5310^{+0.1325}_{-0.1964}$~~~~ & $3.8003^{+0.8178}_{-1.1685}$ \\[0.6mm]

 \hline\hline
 \end{tabular}
 \end{center}
 \vspace{-1mm}  
 \caption{\label{tab5} The same as in Table~\ref{tab4}, but for the
 GPAge parameterizations, by using the observational data SN+FAP+CC
 (SFC), SN+FAP+CC+SH0ES (SFCS), SN+CMB (SC), SN+CMB+FAP (SCF),
 SN+CMB+FAP+CC (SCFC), SN+CMB+FAP+CC+SH0ES (SCFCS).~Note that
 in SGPAge, $H_{0,\,\ssrm LU}$, $\eta_{\ssrm EU}$ and $p_{\rm
 age,\,\ssrm EU}$ should be regarded as $H_0$, $\eta$ and $p_{\rm age}$,
 respectively.~See Sec.~\ref{sec4} for details.}
 \end{sidewaystable}



 \begin{sidewaystable}[tbp] 
 \renewcommand{\arraystretch}{1.7}
 \begin{center}
 \vspace{-1mm}  
 \hspace{1mm}   
 \begin{tabular}{llrrrc|cllrrr}\hline\hline
 Data & ${\cal M}_1-{\cal M}_2$ & $\ln {\cal B}_{12}$ & ~~$\Delta \rm AIC_{12}$ & ~~$\Delta \rm BIC_{12}$ & ~~ & ~~
 & Data & ${\cal M}_1-{\cal M}_2$ & $\ln {\cal B}_{12}$ & ~~$\Delta \rm AIC_{12}$ & ~~$\Delta \rm BIC_{12}$ \\[0.6mm] \hline

 SFC & SGPAge $-$ $\Lambda$CDM & $-2.31$ & $-1.98$ & $8.94$ & & & SFCS & SGPAge $-$ $\Lambda$CDM & $-2.31$ & $-2.05$ & $8.87$ \\
 SFC & TGPAge-Fix $-$ $\Lambda$CDM & $0.15$ & $-2.05$ & $8.88$ & & &SFCS & TGPAge-Fix $-$ $\Lambda$CDM & $0.52$ & $-2.09$ & $8.84$ \\
 SFC & TGPAge-Free $-$ $\Lambda$CDM & $-3.89$ & $0.88$ & $22.72$ & & & SFCS & TGPAge-Free $-$ $\Lambda$CDM & $-3.83$ & $1.08$ & $22.93$ \\
 SFC & TGPAge-Fix $-$ SGPAge & $2.46$ & $-0.07$ & $-0.07$ & & & SFCS & TGPAge-Fix $-$ SGPAge & $2.83$ & $-0.03$ & $-0.03$ \\
 SFC & TGPAge-Free $-$ SGPAge & $-1.59$ & $2.86$ & $13.78$ & & & SFCS & TGPAge-Free $-$ SGPAge & $-1.52$ & $3.13$ & $14.05$ \\
 SFC & TGPAge-Free $-$ TGPAge-Fix & $-4.04$ & $2.92$ & $13.84$ & & & SFCS & TGPAge-Free $-$ TGPAge-Fix & $-4.35$ & $3.16$ & $14.09$ \\[0.6mm] \hline

 SC & SGPAge $-$ $\Lambda$CDM & $29.94$ & $-87.78$ & $-71.46$ & & & SCF & SGPAge $-$ $\Lambda$CDM & $20.06$ & $-68.48$ & $-52.15$ \\
 SC & TGPAge-Fix $-$ $\Lambda$CDM & $33.02$ & $-89.13$ & $-72.81$ & & & SCF & TGPAge-Fix $-$ $\Lambda$CDM & $26.64$ & $-78.73$ & $-62.40$ \\
 SC & TGPAge-Free $-$ $\Lambda$CDM & $29.17$ & $-85.50$ & $-58.30$ & & & SCF & TGPAge-Free $-$ $\Lambda$CDM & $26.83$ & $-86.78$ & $-59.56$ \\
 SC & TGPAge-Fix $-$ SGPAge & $3.08$ & $-1.34$ & $-1.34$ & & & SCF & TGPAge-Fix $-$ SGPAge & $6.58$ & $-10.25$ & $-10.25$ \\
 SC & TGPAge-Free $-$ SGPAge & $-0.78$ & $2.29$ & $13.17$ & & & SCF & TGPAge-Free $-$ SGPAge & $6.77$ & $-18.29$ & $-7.41$ \\
 SC & TGPAge-Free $-$ TGPAge-Fix & $-3.86$ & $3.63$ & $14.51$ & & & SCF & TGPAge-Free $-$ TGPAge-Fix & $0.19$ & $-8.04$ & $2.84$ \\[0.6mm] \hline

 SCFC & SGPAge $-$ $\Lambda$CDM & $19.48$ & $-66.98$ & $-50.59$ & & & SCFCS & SGPAge $-$ $\Lambda$CDM & $20.49$ & $-69.75$ & $-53.36$ \\
 SCFC & TGPAge-Fix $-$ $\Lambda$CDM & $26.16$ & $-77.54$ & $-61.16$ & & & SCFCS & TGPAge-Fix $-$ $\Lambda$CDM & $28.39$ & $-82.36$ & $-65.97$ \\
 SCFC & TGPAge-Free $-$ $\Lambda$CDM & $26.26$ & $-85.10$ & $-57.79$ & & & SCFCS & TGPAge-Free $-$ $\Lambda$CDM & $28.14$ & $-89.92$ & $-62.61$ \\
 SCFC & TGPAge-Fix $-$ SGPAge & $6.68$ & $-10.57$ & $-10.57$ & & & SCFCS & TGPAge-Fix $-$ SGPAge & $7.90$ & $-12.61$ & $-12.61$ \\
 SCFC & TGPAge-Free $-$ SGPAge & $6.78$ & $-18.12$ & $-7.20$ & & & SCFCS & TGPAge-Free $-$ SGPAge & $7.65$ & $-20.17$ & $-9.25$ \\
 SCFC\hspace{5mm} & TGPAge-Free $-$ TGPAge-Fix\hspace{5mm} & $0.10$ & $-7.55$ & $3.37$ & & & SCFCS\hspace{5mm} & TGPAge-Free $-$ TGPAge-Fix\hspace{5mm} & $-0.25$ & $-7.57$ & $3.36$ \\[0.6mm]

 \hline\hline
 \end{tabular}
 \end{center}
 \vspace{-1mm}  
 \caption{\label{tab6} Comparing models ${\cal M}_1$ and ${\cal M}_2$
 by using all the information criteria with the observational
 data SN+FAP+CC (SFC), SN+FAP+CC+SH0ES (SFCS), SN+CMB (SC),
 SN+CMB+FAP (SCF), SN+CMB+FAP+CC (SCFC), SN+CMB+FAP+CC+SH0ES
 (SCFCS).~See Sec.~\ref{sec4} and Table~\ref{tab2} for
 details.}
 \end{sidewaystable}



 \begin{center}
 \begin{figure}[tb]
 \centering
 \vspace{-6.3mm} 
 \includegraphics[width=0.6\textwidth]{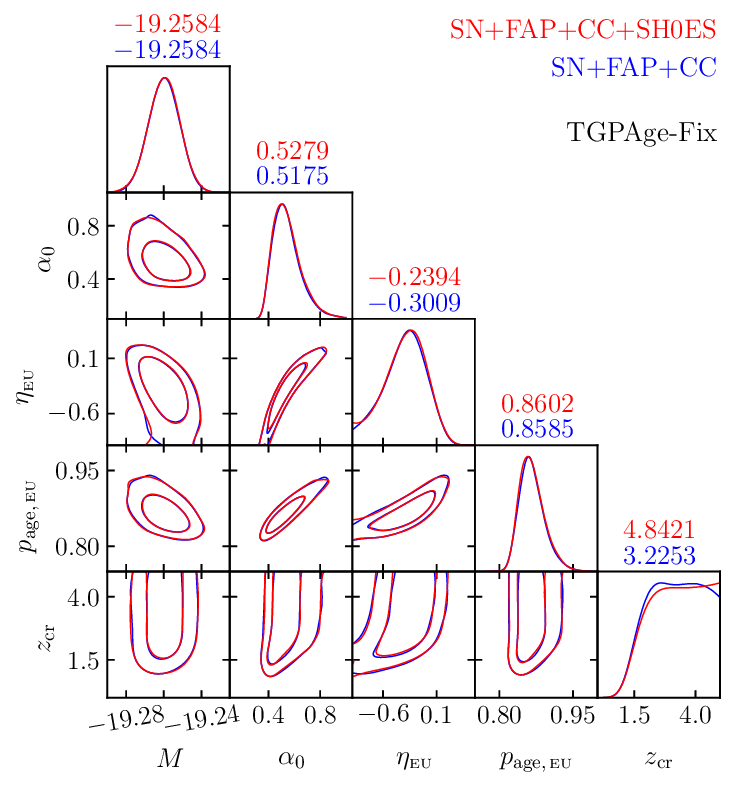}
 \vspace{-1.2mm}  
 \caption{\label{fig4} The same as in Fig.~\ref{fig2}, but for
 TGPAge-Fix.~See Sec.~\ref{sec4} for details.}
 \end{figure}
 \end{center}



 \begin{center}
 \begin{figure}[tb]
 \centering
 \vspace{-6mm}  
 \includegraphics[width=0.85\textwidth]{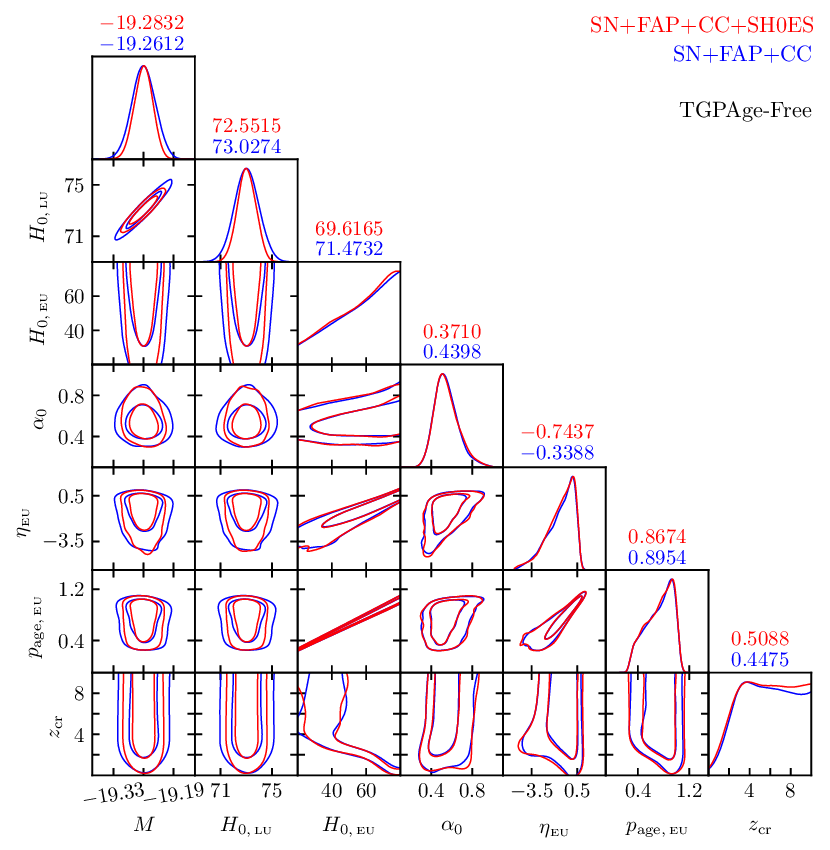}
 \vspace{-1mm}  
 \caption{\label{fig5} The same as in Fig.~\ref{fig2}, but for
 TGPAge-Free.~See Sec.~\ref{sec4} for details.}
 \end{figure}
 \end{center}


\vspace{-19mm} 


\section{Analysis}\label{sec4}

Here, we confront the models with the observational data.~In
 addition to the GPAge parameterizations, we also consider the
 flat $\Lambda$CDM model for comparison, which is given by
 \be{eq34}
 E(z)\equiv H(z)/H_0=\left[\Omega_{m0}\left(1+z\right)^3+\Omega_{r0}
 \left(1+z\right)^4+\left(1-\Omega_{m0}-\Omega_{r0}\right)\right]^{1/2}\,,
 \ee
 where $\Omega_{m0}$, $\Omega_{r0}$ are the present fractional matter
 and radiation densities, respectively.~They are related through the
 redshift of radiation-matter equality $z_{\rm eq}$, namely
 (see e.g.~\cite{Chen:2018dbv})
 \be{eq35}
 \Omega_{r0}=\Omega_{m0}/(1+z_{\rm eq})\,,\hspace{10mm}
 z_{\rm eq}=2.5\times 10^4 \left(T_{\rm CMB}/2.7\,{\rm K}\right)^{-4}
 (\Omega_{m0}h^2)\,,
 \ee
 where $T_{\rm CMB}=2.7255\,{\rm K}$~\cite{Fixsen:2009ug} and
 $h=H_0/(100\;{\rm km/s/Mpc})$.

Since we let $\alpha_0$ be free in all the GPAge parameterizations, at first
 we are interested in whether a free $\alpha_0$ alone could make GPAge much
 better than $\Lambda$CDM by using only the late-time observations without
 CMB.~In this work, we adopt the uniform priors given in Table~\ref{tab3} to
 run MCMC.~In the cases without CMB, $\Omega_{m0}$ is a free parameter
 in the $\Lambda$CDM model.~We fit $\Lambda$CDM, SGPAge, TGPAge-Fix,
 TGPAge-Free to the observational data SN+FAP+CC (SFC) and SN+FAP+CC+SH0ES
 (SFCS), and present the $1\sigma$ and $2\sigma$ constraints on their free
 model parameters in Figs.~$\ref{fig2}-\ref{fig5}$, respectively.~Notice
 that the best-fit values of their model parameters are also given in these
 figures.~Although the marginalized probability distributions and the
 $1\sigma$, $2\sigma$ contours of all the free model parameters have
 been plotted in Figs.~$\ref{fig2}-\ref{fig5}$, we also explicitly give
 their numerical means and $1\sigma$ intervals in Tables~\ref{tab4}
 and \ref{tab5}.~We compare these models by computing $\ln {\cal B}$,
 $\Delta {\rm AIC}$, $\Delta {\rm BIC}$, and present them in the top
 panels of Table~\ref{tab6}.~It is easy to see that $\Delta {\rm BIC}$
 strongly prefers $\Lambda$CDM over all the GPAges, while $\ln {\cal B}$
 and $\Delta {\rm AIC}$ weakly or moderately prefer $\Lambda$CDM over
 two of three GPAges, and the evidences for the remaining one
 are inconclusive.~So, the GPAge parameterizations cannot be much better
 (even worse) than $\Lambda$CDM by using only the late-time observations
 without CMB, in consistent with the no-go
 arguments of~\cite{Cai:2021weh,Cai:2022dkh,Huang:2024erq}.

\begin{center}
 \begin{figure}[tb]
 \centering
 \vspace{-6.4mm} 
 \includegraphics[width=0.45\textwidth]{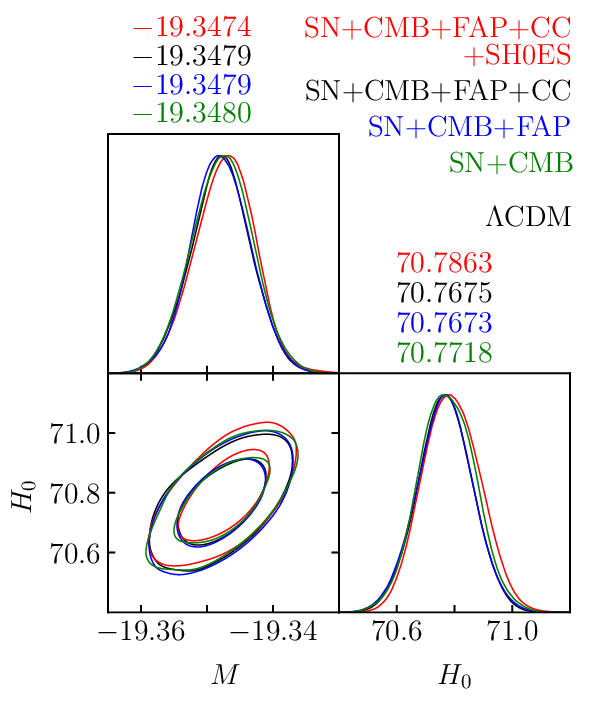}
 \vspace{-2mm}  
 \caption{\label{fig6} The $1\sigma$ and $2\sigma$ contours for all the
 free parameters of the $\Lambda$CDM model from the observational data
 SN +CMB (green), SN+CMB+FAP (blue), SN+CMB+FAP+CC (black),
 SN+CMB+FAP+CC+SH0ES (red).~The marginalized probability distributions
 and the best-fit values are also given at the tops of all columns for
 the corresponding parameters.~The Hubble constant $H_0$ is in
 units of $\rm km/s/Mpc$.~See Sec.~\ref{sec4} for details.}
 \end{figure}
 \end{center}

\begin{center}
 \begin{figure}[tb]
 \centering
 \vspace{-7mm}  
 \includegraphics[width=0.6\textwidth]{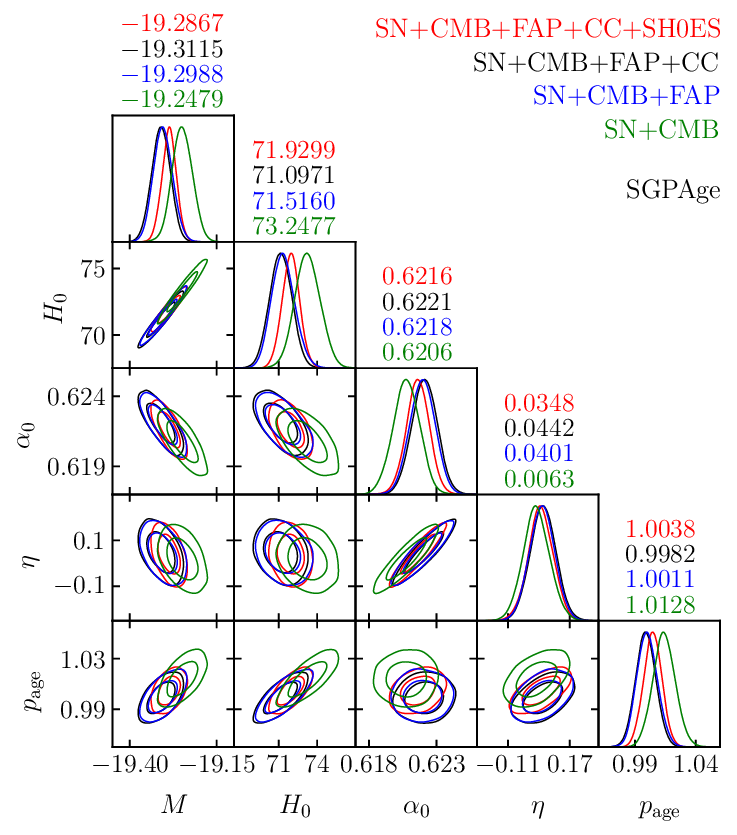}
 \vspace{-2mm}  
 \caption{\label{fig7} The same as in Fig.~\ref{fig6}, but for
 SGPAge.~See Sec.~\ref{sec4} for details.}
 \end{figure}
 \end{center}

\begin{center}
 \begin{figure}[tb]
 \centering
 \vspace{-7mm}  
 \includegraphics[width=0.6\textwidth]{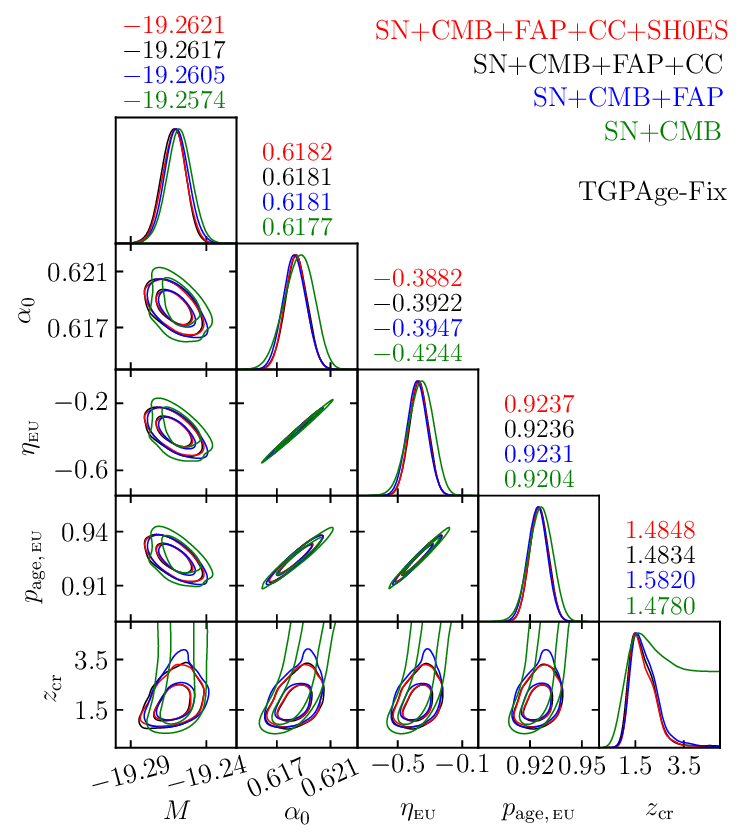}
 \vspace{-2mm}  
 \caption{\label{fig8} The same as in Fig.~\ref{fig6}, but for
 TGPAge-Fix.~See Sec.~\ref{sec4} for details.}
 \end{figure}
 \end{center}

\begin{center}
 \begin{figure}[tb]
 \centering
 \vspace{-6.4mm} 
 \includegraphics[width=0.85\textwidth]{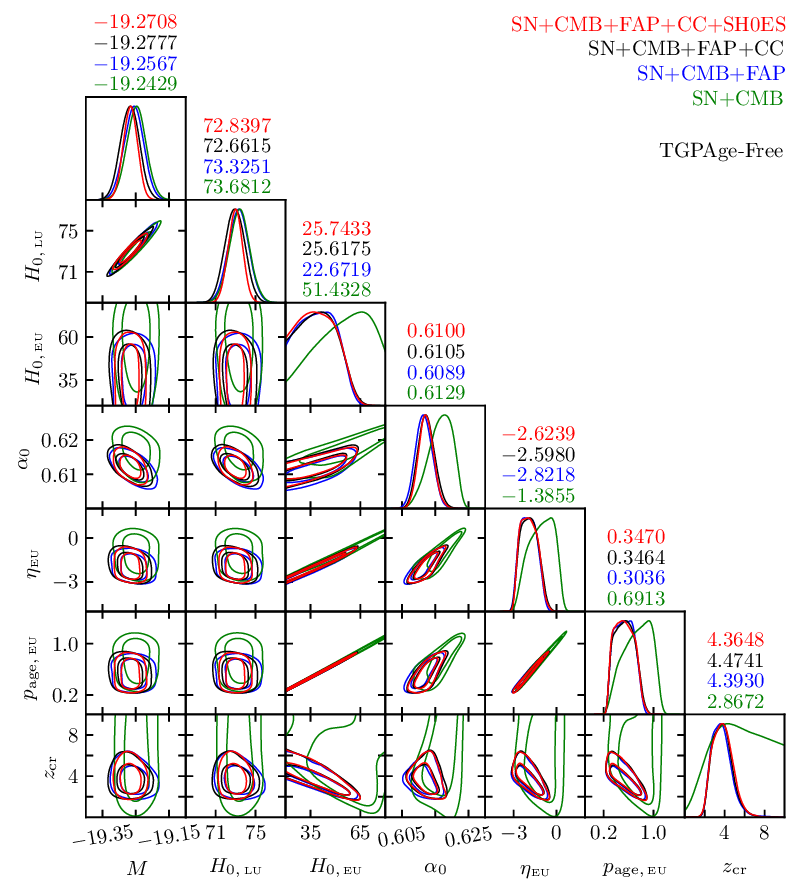}
 \vspace{-1.5mm}  
 \caption{\label{fig9} The same as in Fig.~\ref{fig6}, but for
 TGPAge-Free.~See Sec.~\ref{sec4} for details.}
 \end{figure}
 \end{center}

Fortunately, the GPAge parameterizations have been extended to redshift
 $z>z_{\rm CMB}$, and hence the CMB observational data can be also taken
 into account.~Note that in the cases with CMB, $\Omega_{m0}$ is no longer a
 free parameter in the $\Lambda$CDM model, due to the discussions around
 Eqs.~(\ref{eq26}) and (\ref{eq27}).~We fit $\Lambda$CDM,
 SGPAge, TGPAge-Fix, TGPAge-Free to the observational data SN+CMB (SC),
 SN+CMB+FAP (SCF), SN+CMB+FAP+CC (SCFC), SN+CMB+FAP+CC+SH0ES (SCFCS) one
 by one, and present the $1\sigma$ and $2\sigma$ constraints and the
 best-fit values of their free model parameters in Figs.~$\ref{fig6}-
 \ref{fig9}$ respectively, while their numerical means and
 $1\sigma$ intervals are also given in Tables~\ref{tab4} and
 \ref{tab5}.~We compare these models by computing $\ln {\cal B}$,
 $\Delta {\rm AIC}$, $\Delta {\rm BIC}$, and present them in the middle
 and bottom panels of Table~\ref{tab6}.

From the middle-left panel of Table~\ref{tab6}, using the observational
 data SN+CMB (SC), all the $\ln {\cal B}$, $\Delta {\rm AIC}$, $\Delta
 {\rm BIC}$ overwhelmingly prefer all the three GPAge parameterizations
 over the $\Lambda$CDM model.~This is an impressive result.~On the other
 hand, TGPAge-Fix and TGPAge-Free are not much better (even worse) than
 SGPAge by using only the SN+CMB (SC) data.

Adding the observational data of $F_{\rm AP}$ and using the SN+CMB+FAP
 (SCF) data, it is easy to see from the middle-right panel of
 Table~\ref{tab6} that all the three GPAge parameterizations are still
 overwhelmingly preferred over the $\Lambda$CDM model, and now both
 TGPAge-Fix and TGPAge-Free have come to be strongly preferred over
 SGPAge.~The observational data of $F_{\rm AP}$ make difference.~But in
 this case, TGPAge-Free is not much better (even worse) than TGPAge-Fix.

We further add the observational data CC and SH0ES one by one.~From the
 bottom panels of Table~\ref{tab6}, we find that although the
 evidences of $\ln {\cal B}$, $\Delta {\rm AIC}$, $\Delta {\rm BIC}$
 become slightly stronger, the main results are not so different from
 the case of SN+CMB+FAP (SCF).

Since the Hubble tension is mainly between CMB and SNIa/Cepheids, it is
 necessary to simultaneously consider both CMB and SNIa at least.~As
 shown above, once CMB is taken into account, the situation has been
 significantly changed.~Contrary to the no-go arguments
 of~\cite{Cai:2021weh,Cai:2022dkh,Huang:2024erq}, we find that all the
 GPAge parameterizations are overwhelmingly preferred over the standard
 $\Lambda$CDM model in terms of all the information criteria $\ln {\cal
 B}$, $\Delta {\rm AIC}$, $\Delta {\rm BIC}$.~There should be new physics in
 the difference between GPAges and $\Lambda$CDM, which will be further
 discussed in the next section.~In addition, both TGPAges are strongly
 preferred over SGPAge. In TGPAge, as mentioned in Sec.~\ref{sec2b},
 $H_{0,\,{\ssrm EU}}$ and $H_{0,\,{\ssrm LU}}$ correspond to the Hubble
 constant inferred from CMB and the Hubble constant measured by SH0ES
 with nearby Cepheids/SNIa, respectively.~Note that in
 TGPAge-Free, $H_{0,\,{\ssrm EU}}$ cannot be well constrained
 and its best fit slightly deviates from $67.36\pm 0.54\;{\rm km/s/Mpc}$
 inferred from Planck 2018~\cite{Planck:2018vyg} (while its
 $H_{0,\,{\ssrm LU}}$ is in well consistent with $73.04\pm 1.04\;{\rm
 km/s/Mpc}$ directly measured by SH0ES), and only $\Delta\rm AIC$ weakly
 prefers TGPAge-Free over TGPAge-Fix while both $\Delta\rm BIC$ and
 $\ln {\cal B}$ oppose.~So, it is reasonable to regard TGPAge-Fix as the
 best, in which the Hubble tension could be naturally alleviated (or
 even resolved).


 \begin{center}
 \begin{figure}[tb]
 \centering
 \vspace{-7mm}  
 \hspace{-3mm}  
 \includegraphics[width=0.98\textwidth]{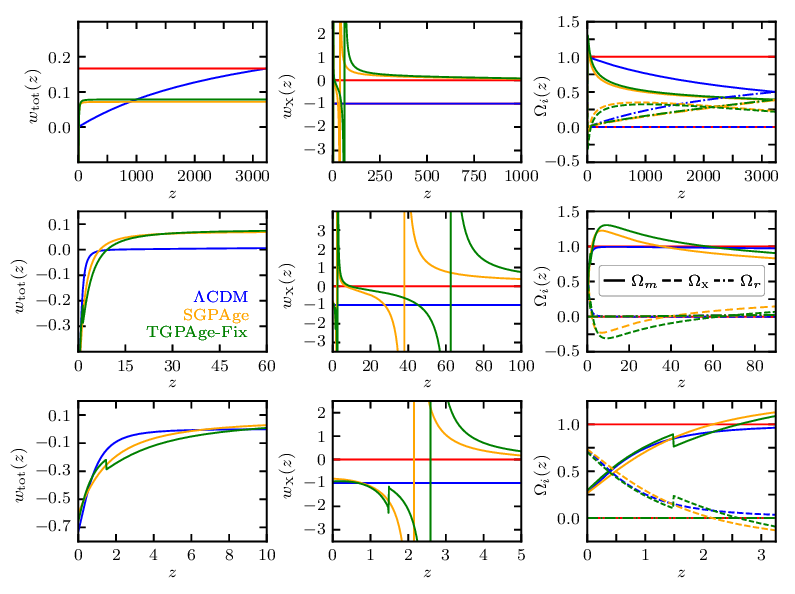}
 \vspace{-1.5mm}  
 \caption{\label{fig10} $w_{\rm tot}$ (left), $w_{\ssrm X}$ (middle) and
 $\Omega_i$ (right) as functions of redshift $z$ for the $\Lambda$CDM
 (blue), SGPAge (yellow) and TGPAge-Fix (green) models with the best-fit
 parameters from the observational data SN+CMB+FAP+CC (SCFC).~The top
 panels are plotted for the entire cosmic history $0\leq z\leq z_{\rm
 eq}$, while the middle and bottom panels have been zoomed at middle
 and low redshifts to show details, respectively.~In the right panels,
 $\Omega_m$, $\Omega_r$ and $\Omega_{\ssrm X}$ are plotted as solid,
 dash-dot and dashed lines, respectively.~The critical values $\Omega_i
 =0$ and $1$, $w_{\ssrm X}=0$ and $-1$, $w_{\rm tot}=1/6$ are also indicated
 by the red solid horizontal lines. See Sec.~\ref{sec5} for details.}
 \end{figure}
 \end{center}


\vspace{-41mm} 


\section{Physical implications}\label{sec5}

So far, with the GPAge parameterizations, we are completely ignorant of
 the components in the universe and the gravity theory.~As shown in
 Sec.~\ref{sec2}, we even do not know whether the universe contains dark
 energy, dark matter or other components, and in fact we even have not
 used any Friedmann equation or Einstein equation so far, while
 no particular metric or geometry is explicitly assumed.~Thus,
 the results obtained in the above sections are fairly general.~We have
 found that the Hubble tension can be alleviated (or even resolved) in
 GPAges different from the standard $\Lambda$CDM cosmology, while they
 are overwhelmingly preferred over $\Lambda$CDM by the observational
 data in terms of all the information criteria.

We cannot satisfy and just stop here.~In this section, we try to look
 deeper into the new physics behind the GPAge parameterizations.~Let us
 go further with some solid reality-motivated assumptions.~In our real
 universe, there certainly exist pressureless (dust) matter
 and radiation, which build our real world and appear in our everyday
 life.~If there are other unknown components in the universe and the
 {\em effective components} from the possible modifications of
 gravity/geometry, we call them ``\;X\,'' in
 a collective.~Thus, the (first) Friedmann equation is given by
 \be{eq36}
 H^2=\frac{8\pi G}{3}\rho_{\rm tot}=
 \frac{8\pi G}{3}\left(\rho_m+\rho_r+\rho_{\ssrm X}\right)\,.
 \ee
 Although it apparently has the same form as in general relativity (GR),
 one can also obtain it in modified gravity by incorporating all the
 modifications to GR into the {\em effective} energy components
 $\rho_{\ssrm X}\,$.~Similarly, the Raychaudhuri equation (the second
 Friedmann equation) is given by
 \be{eq37}
 \dot{H}=-4\pi G\left(\rho_{\rm tot}+p_{\rm tot}\right)\,,
 \ee
 where a dot denotes a derivative with respect to $t$.~Using
 Eqs.~(\ref{eq36}) and (\ref{eq37}), we obtain
 \be{eq38}
 w_{\rm tot}=\frac{p_{\rm tot}}{\rho_{\rm tot}}=-1-\frac{2\dot{H}}{3H^2}
 =-1+\frac{2\left(1+z\right)}{3H}\frac{dH}{dz}\,,
 \ee
 as a function of redshift $z$.~On the other hand, by definition, it is
 well known that
 \be{eq39}
 w_{\rm tot}=\sum_i w_i\Omega_i=w_{\ssrm X}\Omega_{\ssrm X}+\Omega_r/3\,,
 \ee
 where $w_i=\rho_i/p_i$ is the equation-of-state parameter,
 $\Omega_i=\rho_i/\rho_{\rm tot}$ is the fractional energy density,
 and we have used $w_m=0$ and $w_r=1/3$.~Using Eq.~(\ref{eq39}) and
 \be{eq40}
 \Omega_{\ssrm X}(z)=1-\Omega_m(z)-\Omega_r(z)\,,\quad
 \Omega_m(z)=\Omega_{m0}\left(1+z\right)^3/E^2(z)\,,\quad
 \Omega_r(z)=\Omega_{r0}\left(1+z\right)^4/E^2(z)\,,
 \ee
 it is easy to find $w_{\ssrm X}(z)$ as a function of redshift $z$, namely
 \be{eq41}
 w_{\ssrm X}=\left(w_{\rm tot}-\Omega_r/3\right)/\Omega_{\ssrm X}\,,
 \ee
 where $w_{\rm tot}$ is given by Eq.~(\ref{eq38}), and $\Omega_{r0}$ is
 given by Eq.~(\ref{eq35}).
 \vspace{-1mm}\newpage 

In Fig.~\ref{fig10}, we plot $w_{\rm tot}$, $w_{\ssrm X}$ and $\Omega_i$
 as functions of redshift $z$ for the $\Lambda$CDM, SGPAge and
 TGPAge-Fix models with the best-fit parameters from the observational
 data SN+CMB+FAP+CC (SCFC).~The top panels are plotted for the entire
 cosmic history $0\leq z\leq z_{\rm eq}$, while the middle and bottom
 panels have been zoomed at middle and low redshifts to show details,
 respectively.~From Fig.~\ref{fig10}, it is easy to see that
 $\Lambda$CDM behaves as well known, $w_{\rm tot}\to 1/6=(1/3)
 \cdot (1/2)$ when $z\to z_{\rm eq}$ where $\Omega_r\simeq\Omega_m$ and
 $\Omega_\Lambda\simeq 0$, while $\Lambda$ makes sense only at fairly
 low redshifts.~However, GPAges significantly deviate from $\Lambda$CDM
 at middle and high redshifts.~In GPAges, $w_{\ssrm X}\to 0$ at very
 high redshifts (n.b.~the top-middle pannel of Fig.~\ref{fig10}), and
 $w_{\ssrm X}$ diverges twice around $z\sim 40-60$ and $z\sim 2-3$ (n.b.~all
 the $w_{\ssrm X}$ panels of Fig.~\ref{fig10}) where $\Omega_{\ssrm X}$
 crosses $0$ twice (n.b.~the middle-right and bottom-right panels of
 Fig.~\ref{fig10} and Eq.~(\ref{eq41})).~It is worth noting
 that $\Omega_{\ssrm X}<0$ (and hence $\Omega_m>1$) at middle redshifts
 $2\lesssim z\lesssim 60$ (n.b.~the middle-right panel of
 Fig.~\ref{fig10}).~As is well known, any real energy component in the
 universe (including dark energy or dark matter) requires $\rho_i\geq 0$
 or equivalently $\Omega_i\geq 0$.~Only the {\em effective components}
 from the modifications of gravity/geometry could have ({\em effective})
 negative $\Omega_{\ssrm X}$ without any problem.~So, at least
 in the mid-time, the gravity/geometry must be significantly modified
 (n.b.~$\Omega_{\ssrm X}$ can be smaller than $-0.25\sim -0.3$ at most,
 see the middle-right panel of Fig.~\ref{fig10}).~As is well known,
 modified gravity and interacting dark energy
 cannot be distinguished~\cite{Kunz:2006ca,Wei:2008vw,Bertschinger:2008zb,
 Wei:2013rea}.~So, at least in the mid-time (say, $2\lesssim z\lesssim 60$),
 modified gravity and/or interacting dark energy should be invoked,
 while all the non-interacting dark energy models in GR could
 be excluded.~The new physics in the mid-time might be the key
 to resolve the Hubble tension.

In fact, the possible need of $\rho_{\ssrm DE}<0$ around $z\gtrsim 1.6$
 was first highlighted by the BAO data of BOSS~\cite{BOSS:2014hhw} (see
 also e.g.~Sec.~4.2.2 of~\cite{DiValentino:2025sru}).~Recently, the DESI
 BAO data also hint $\rho_{\ssrm DE}<0$ at
 $z\gtrsim 1.5-2$~\cite{DESI:2024aqx,Escamilla:2024ahl}. Actually, they
 saw nothing but $\Omega_{\ssrm X}<0$ in the right corner of
 the bottom-right panel of our Fig.~\ref{fig10}, namely the tip of the
 iceberg of the mid-time (say, $2\lesssim z\lesssim 60-100$).~The whole
 elephant is shown by the middle-right panel (and the top-right
 panel) of our Fig.~\ref{fig10}.

Let us turn to $w_{\ssrm X}$.~As mentioned above, $w_{\ssrm X}\to 0$ at
 very high redshifts, and hence it plays the role of pressureless (dust)
 matter in the early time (where $\Omega_{\ssrm X}\sim 0.25$ cannot be
 ignored, n.b.~the top-right panel of Fig.~\ref{fig10}).~In all the
 $w_{\ssrm X}$ panels of Fig.~\ref{fig10}, $w_{\ssrm X}$ diverges twice
 due to $\Omega_{\ssrm X}$ crossing $0$ twice (n.b.~Eq.~(\ref{eq41})),
 as mentioned above.~In between, $w_{\ssrm X}$ crosses $-1$ and $0$ from
 below to above (n.b.~the middle-middle panel of Fig.~\ref{fig10}).~Note
 that $\Omega_{\ssrm X}$, $w_{\ssrm X}$ and $w_{\rm tot}$ of TGPAge jump
 once at $z_{\rm cr}$ (n.b.~the bottom panels of Fig.~\ref{fig10}),
 since $E(z)$ is not continuous at $z_{\rm cr}$, as mentioned
 in Sec.~\ref{sec2b}.~At low redshifts, $w_{\ssrm X}$ crosses $-1$ from
 $w_{\ssrm X}<-1$ to $w_{\ssrm X}>-1$ (n.b.~the bottom-middle panel of
 Fig.~\ref{fig10}).~Recently, the DESI BAO data prefer a dynamical dark
 energy with $-1<w_0<0$, $w_a<0$ and $w_0+w_a<-1$~\cite{DESI:2025zgx},
 namely $w_{\ssrm DE}$ crosses $-1$ from below to above, in consistent
 with our result.

Finally, from the bottom panels of Fig.~\ref{fig10}, it is easy to see
 that GPAges are fairly close to $\Lambda$CDM at low redshifts
 $z\lesssim 1.5$, especially $z\lesssim 1$.~So, there is only little
 room for the modifications beyond $\Lambda$CDM in the late
 time ($z\lesssim 1.5$, especially $z\lesssim 1$), partly in consistent
 with~\cite{Cai:2021weh,Cai:2022dkh,Huang:2024erq}.~The key of
 the present work is in the mid-time (say, $2\lesssim z\lesssim
 60-100$), rather than the late or the early time.~We show that the new
 physics in the mid-time makes big differences.


\section{Concluding remarks}\label{sec6}

In cosmology, the Hubble tension has become one of the most serious
 challenges in the last decade. Assuming that all the observational
 data are right, one of the ways out is to modify the standard
 $\Lambda$CDM cosmology. In the literature, the theoretical
 modifications are mainly made in the early and the late/local
 universes.~But on both sides, some no-go arguments have been
 proposed.~In particular, the general no-go arguments for the
 late-time modifications in~\cite{Cai:2021weh,Cai:2022dkh,
 Huang:2024erq} are mainly based on the PAge parameterization.~In the
 present work, we note that there are some flaws in the (original) PAge
 parameterization, and then propose various generalized PAge (GPAge)
 parameterizations, which can be extended to $z>z_{\rm CMB}$ and hence
 CMB can be taken into account, while they are also more accurate at low
 redshifts where $\tau=t/t_0$ is close to $1$.~With these GPAge
 parameterizations, we revisit the Hubble tension by using not only the
 late-time observations but also the observational data of CMB.~We find
 that the Hubble tension could be alleviated (or even resolved)
 in GPAges different from the standard $\Lambda$CDM cosmology, while
 they are overwhelmingly preferred over $\Lambda$CDM by the
 observational data in terms of all the information criteria.
 Surprisingly, we find that the new physics in the mid-time might be the
 key to resolve the Hubble tension.

Some remarks are in order.~First, our results strongly prefer modified
 gravity (while interacting dark energy and modified gravity cannot
 be distinguished~\cite{Kunz:2006ca,Wei:2008vw,Bertschinger:2008zb,
 Wei:2013rea}), and all the non-interacting dark energy models in GR
 can be excluded.~However, noting that only the observational data of
 the cosmic expansion history have been used in the present work, this
 conclusion is in fact not so solid without also using the observational
 data of the growth history.~As is well known, modified gravity affect
 both the expansion history and the growth history in a way different
 from non-interacting dark energy.~If the non-interacting dark energy
 model and the modified gravity model share the same cosmic expansion
 history, they might have different growth histories~\cite{Kunz:2006ca,
 Wei:2008vw,Bertschinger:2008zb,Wei:2013rea}.~Thus, it is needed to also
 take the observational data of the growth history (e.g.~the well-known
 $f\sigma_8$ data) into account.~Without them, our key conclusions are
 not so solid in fact.~We will continue to study the mid-time guide for
 the Hubble tension with the $f\sigma_8$ data and/or other
 observational data of the growth history in our next works.

Second, in e.g.~\cite{Yin:2018mvu}, the non-parametric reconstruction of
 growth index by using the observational data of $f\sigma_8$, SNIa and
 CC was considered.~The reconstructed growth index strongly prefer a
 modified gravity scenario different from $f(R)$ theories, while $f(R)$
 theories and non-interacting dark energy models in GR are strongly
 disfavored.~In fact, the reconstructed $\Omega_m(z)$ given by
 Figs.~$3-6$ of~\cite{Yin:2018mvu} becomes $\Omega_m>1$ at redshifts
 $z\gtrsim 1.5-2$, in well consistent with our $\Omega_m(z)$ of the
 GPAge parameterizations (n.b.~the middle-right and the bottom-right
 panels of our Fig.~\ref{fig10}).~Noting that the $f\sigma_8$ data of
 the growth history has also been used in~\cite{Yin:2018mvu},
 the results of~\cite{Yin:2018mvu} might be regarded as a support for
 the mid-time guide for the Hubble tension proposed in the present work.

Third, in this work, we only use the distance priors from the
 CMB observations.~In fact, it is much better to use the full CMB data
 of Planck 2018 instead.~To this end, a modified CAMB or CLASS should
 be used with Cobaya or CosmoMC.~This is not so easy but deserves to do
 in the future works.

Fourth, most of the known cosmological probes are in the late time, such
 as SNIa, CC, BAO, FRBs. Only a few data points of them are at redshifts
 $z>2$ (for instance, only one data point in the Pantheon+ SNIa sample
 is at $z>2$).~On the other hand, the CMB observation is at
 high redshift $z\sim 1090$.~However, as shown in the present work, the
 new physics in the mid-time (say, $2\lesssim z\lesssim 60-100$) is the
 key to resolve the Hubble tension.~So, the observations in the mid-time
 are valuable.~To our knowledge, the redshifts of gamma-ray
 bursts (GRBs) and quasars can be up to $z\sim 10$.~They could be calibrated
 as standard candles, and hence we might use them to test the mid-time guide
 for the Hubble tension proposed here. We leave them to our next works.

Fifth, the GPAge parameterizations proposed in this work might
 be further improved.~For example, the TGPAge parameterizations are not
 so smooth at the critical redshift $z_{\rm cr}$, and hence we find that
 their $\Omega_i$, $w_{\rm tot}$ and $w_{\ssrm X}$ jump once at $z_{\rm
 cr}$, as shown in Fig.~\ref{fig10}.~Thus, it is desirable to find some
 well-behaved new GPAge parameterizations in the future works.~On the
 other hand, the GPAge parameterizations proposed in this work could be
 enough accurate at both low redshifts and high redshifts, but at the
 middle redshifts their accuracies have not been well guaranteed in
 principle.~To study the new physics in the mid-time, namely the key to
 resolve the Hubble tension, more accurate GPAge parameterizations in
 the mid-time are also needed.~In fact, they are not necessary to be
 a new kind of PAge parameterizations, any well-motivated
 non-PAge parameterizations to this end are also desirable.

Finally, as mentioned above, our results strongly prefer
 modified gravity.~As is well known, gravity is a geometric effect.~So,
 if the geometry is modified, for example, if there is a giant
 underdense (void) or overdense region nearby us and hence the universe
 should be characterized by the Lema\^{i}tre-Tolman-Bondi (LTB) metric
 (see e.g.~\cite{Jia:2025prq} and the references therein) rather than
 the well-known Friedmann-Robertson-Walker (FRW) metric, it
 is also viable.~Noting Eq.~(\ref{eq40}) and $E(z)$ is not continuous
 at redshift $z_{\rm cr}$ in TGPAge as mentioned in Sec.~\ref{sec2b},
 $\Omega_m(z)$ and then $\Omega_{\ssrm X}(z)=1-\Omega_m(z)-
 \Omega_r(z)$ are discontinuous at $z_{\rm cr}$ (n.b.~the bottom-right
 panel of Fig.~\ref{fig10}).~In the framework of a giant underdense
 or overdense region nearby us, it is easy to understand the sudden
 transition of $\Omega_m$ around the edge of this region.~From
 the bottom-right panel of Fig.~\ref{fig10}, we can find that $\Omega_m
 (z_{\rm cr})$ inside this region is larger than $\Omega_m(z_{\rm cr})$
 outside this region.~So, it is in fact an overdense region, rather than
 a void.~On the other hand, the best-fit $z_{\rm cr}\sim 1.5$ in the
 TGPAge-Fix parameterization by using the observational data
 SN+CMB+FAP+CC (SCFC), as shown in Figs.~\ref{fig10} and
 \ref{fig8}.~Thus, it is in fact a cosmological (rather than local)
 overdense region.~This is another surprise, since in the literature a
 local void was commonly used instead.~Our results suggest that
 a cosmological overdense region might be a new solution for the Hubble
 tension.~However, we stress that such a cosmological overdense region
 is significantly different from the local one, since its $\Omega_m(z)$
 inside this region ($z<z_{\rm cr}$) becomes larger when $z$ increases
 (n.b.~the bottom-right panel of Fig.~\ref{fig10}), namely $\Omega_m$
 is dependent on the cosmic time $t$ (equivalent to redshit $z$), while
 $\Omega_m$ is time-independent in a local underdense/overdense
 region (mainly due to its locality, see e.g.~\cite{Jia:2025prq} and the
 references therein).~So, we might find something new to resolve the
 Hubble tension.~This is an interesting new direction deserving
 further study in the future works.

\vspace{-2.4mm} 


\section*{ACKNOWLEDGEMENTS}

We are grateful to Profs.~Shao-Jiang~Wang and Yong~Zhou, as well as
 Yuxuan~Shi, Shu-Yan~Long, Hui-Qiang~Liu, Wei-Zhi~Gong for kind
 help and useful discussions.~This work was supported in part by NSFC
 under Grants No.~12375042 and No.~11975046.~Da-Chun~Qiang was supported
 in part by NSFC under Grant No.~12505070, the Henan Provincial Natural
 Science Foundation No.~252300420902, and the Startup Research Fund
 of Henan Academy of Sciences No.~241841222.

\renewcommand{\baselinestretch}{1.1}


\end{document}